\documentclass[preprint]{aastex}

\usepackage{natbib}
\usepackage{txfonts}
\usepackage{hyperref}

\bibpunct[; ]{(}{)}{,}{a}{}{;}

\shorttitle{Solar Flare Impulsive Phase Emission}
\shortauthors{Kennedy et al.}

\begin{document}

\title{Solar Flare Impulsive Phase Emission Observed with SDO/EVE}
\author{Michael B. Kennedy\altaffilmark{1}, Ryan O. Milligan\altaffilmark{1,2,3}, Mihalis Mathioudakis\altaffilmark{1} and Francis P. Keenan\altaffilmark{1}}
\altaffiltext{1}{Astrophysics Research Centre, School of Mathematics \& Physics, Queen's University Belfast, University Road, Belfast, UK, BT7 1NN}
\altaffiltext{2}{Solar Physics Laboratory (Code 671), Heliophysics Science Division, NASA Goddard Space Flight Center, Greenbelt, MD 20771, USA}
\altaffiltext{3}{Department of Physics, Catholic University of America, 620 Michigan Ave., N.E., Washington, DC 20064}
\email{mkennedy29@qub.ac.uk}

\begin{abstract}
\noindent
Differential emission measures (DEMs) during the impulsive phase of solar flares were constructed using observations from the EUV Variability Experiment (EVE) and the Markov-Chain Monte Carlo method. Emission lines from ions formed over the temperature range log T$_{e}$ = 5.8 - 7.2 allow the evolution of the DEM to be studied over a wide temperature range at 10s cadence. The technique was applied to several M- and X-class flares, where impulsive phase EUV emission is observable in the disk-integrated EVE spectra from emission lines formed up to 3 - 4 MK, and we use spatially-unresolved EVE observations to infer the thermal structure of the emitting region. For the nine events studied the DEMs exhibited a two component distribution during the impulsive phase, a low temperature component with peak temperature of 1 - 2 MK, and a broad high temperature one from  7 - 30 MK. A bimodal high temperature component is also found for several events, with peaks at 8 and 25 MK during the impulsive phase. The origin of the emission was verified using AIA images to be the flare ribbons and footpoints, indicating that the constructed DEMs represent the spatially-average thermal structure of the chromospheric flare emission during the impulsive phase.
\end{abstract}


\section{Introduction}\label{s:intro}

In the standard model of solar flares, energy is deposited in the lower layers of the atmosphere by non-thermal electron beams which undergo Coulomb collisions when they encounter the dense chromospheric plasma. Hard X-rays are produced through thick-target bremsstrahlung radiation \citep{brow71} and impulsive emission from the flare footpoints is observable in soft X-ray, EUV, UV and white light radiation. The impulsive response in the EUV was first inferred over 40 years ago \citep{donn69} using Sudden Frequency Deviations, sensitive to the broadband solar EUV irradiance in the 1 - 103 nm range. The impulsive EUV bursts were found to be correlated with hard X-ray emission \citep{kane71,kane78}, which provided evidence that non-thermal electrons were responsible for transferring energy to the lower atmosphere. The \textit{Yohkoh} Soft X-ray Telescope (SXT) provided imaging which showed that the impulsive phase emission was observable down to soft X-ray wavelengths \citep{tier93, huds94}, indicating material at the flare footpoints was heated to temperatures of up to 10 MK. Further studies into the connection between footpoint soft and hard X-ray emission indicate that low energy non-thermal electrons directly heat the top of the dense lower atmosphere producing the impulsive soft X-ray emission \citep{mroz04, mroz07}. 

The radiative losses from an optically thin plasma can be described by the differential emission measure (DEM) which indicates the amount of emitting material as a function of temperature. It is an observational quantity that potentially can be useful in determining flare heating mechanisms. Theoretical and semi-empirical models of flare heating show that the energy balance of the flaring atmosphere is dominated by thermal conduction and radiation \citep{mach79,mach80}. Models that make different assumptions about the heating function and energy balance can produce DEMs with different forms, e.g. \citet{emsl85}.

The majority of previous flare DEM studies have focused on the coronal loop emissions, and the only large-scale survey of flare DEMs in the literature is that of \citet{tier99}. Using data from the \textit{Yohkoh} SXT and BCS, a Maximum-Entropy Method was employed to obtain the DEM for 80 flares ranging from high C to low M-class. The flare DEM has been analysed for several long duration X-class flares by \citet{warr13} using the \textit{Extreme-ultraviolet Variability Experiment} (EVE) instrument \citep{wood12}. Assuming a parameterised form for the DEM, a forward-fitting method was adopted to obtain synthetic spectra that best-fit EVE observations, demonstrating the capability of EVE to measure thermal flare plasma over a wide temperature range.

Spatially resolved spectroscopic observations of the EUV emission from flare footpoints during the impulsive phase are rare, due to the requirement of needing the spectrometer slit placed over the correct location on the solar disk at the beginning of the flare. In a recent study \citet{grah13} were able to obtain the impulsive phase emission measure distributions (EMDs) of flare footpoints for the first time using \textit{Hinode}/EIS \citep{culh07}. The EMDs obtained for a number of B and low C-class events showed chromospheric material at the footpoints heated to temperatures of up to log T$_{e}$ = 6.9. A consistent emission measure gradient of EM(T$_{e}$) $\propto$ T$^{1}_{e}$ was found for all events studied, which the authors interpret as a scenario where energy deposited by non-thermal electrons heats the top of the flare chromosphere, with lower layers of the atmosphere heated by conduction.

In this study we present observations of impulsive phase emission made by the EVE instrument onboard the \textit{Solar Dynamics Observatory}. The high cadence and almost continuous observations make EVE ideal for studying the evolution of solar flare plasmas. A DEM analysis is used to determine the spatially-averaged thermal structure of the observed emission, and imaging from the AIA instrument is employed to verify that the emission observed by EVE is originating from the flare ribbons and footpoints during the impulsive phase. In Section \ref{s:obs} the observations and analysis of EVE flare lightcurves are discussed, while the DEM construction technique is described in Section \ref{s:dem}. The results of the DEM analysis are in Section \ref{s:results}, discussion and interpretation in Section \ref{s:dis} and the conclusions are presented in Section \ref{s:conclu}.

\section{Observations}\label{s:obs}

The EVE MEGS-A instrument obtains Sun-as-a-star observations over the 6-37 nm wavelength range at a cadence of 10s and a spectral resolution of 0.1nm. This wavelength range includes emission lines from ions formed over a wide temperature range, and the high cadence and nearly 100\% duty cycle allow the study of the evolution of many solar flares as a function of temperature throughout the entire flare duration \citep{cham12}. The EVE Level 2 Version 3 data release is used in this study as the line irradiance is in better agreement with theory compared to Version 2 \citep{zann13}.

In EVE observations the response of the lower atmosphere to heating and ionisation is easily observable as increased emission from chromospheric and transition region lines such as the \ion{He}{2} 30.4nm doublet. Impulsive phase emission can also be observed from ions formed up to temperatures of $\approx$ 0.5 - 3 MK (e.g. \ion{Fe}{8} to \ion{Fe}{16}) in the disk-integrated EVE spectra. The irradiance increases to a peak co-temporal with the \ion{He}{2} emission, then declines before reaching a second emission peak in the gradual phase as the coronal loops cool. Impulsive high-temperature emission from $\approx$ 10 MK plasma is observable in EUV and SXR images of flare footpoints, but in EVE spectra  does not exhibit an impulsive time profile as it is indistinguishable from the coronal loop emission. 

An analysis of the EVE flare lightcurves was made with the aim of detecting flares where the impulsive phase increase is sufficient to be detected above the pre-flare background emission and can be temporally isolated from any emission originating from the coronal loops in the gradual phase. Then, using the temperature coverage and high cadence of EVE, flare DEMs can be constructed over a wide temperature range, and as a function of time. The aim of this study is to attempt to derive the thermal structure of the impulsive phase flare footpoints and ribbons from spatially-unresolved EVE observations.

\subsection{Flare Sample Selection}

A method was defined to detect the strong impulsive phase emission in lightcurves created from EVE spectral line fluxes, which would provide a sample of flares to be investigated in the DEM analysis. It was based on the EVE temperature evolution code from \citet{cham12} and was used to study the EUV lightcurves for a large flare sample; all GOES flares of C6 or higher from the beginning of EVE science operations on 2010 May 1 until 2012 October 31. This provided an initial sample of 455 flares. There were 38 flares rejected due to gaps in observations preventing accurate identification of the impulsive phase peak, and a further 27 flares were excluded due to overlap in emission from previous events, or if it appeared that multiple flares were counted as a single GOES event. 

For each of the remaining 390 flares the following analysis was performed. A pre-flare average irradiance was subtracted from each EVE wavelength bin to obtain the flare spectra. The irradiance of the strongest, isothermal emission lines formed over a temperature range of log T$_{e}$ = 5.8 - 7.2 in the 6 - 37 nm wavelength range (see Table \ref{t:linelist}) were measured over the entire flare duration. Emission lines were then allocated into five temperature bins separated by 0.4 dex. The \ion{He}{2} 30.4nm doublet was also used to provide a temperature point at log T$_{e}$ = 5.0 and the peak of \ion{He}{2} emission was employed as a proxy for the peak of the impulsive phase. The detection of the impulsive emission was made for each flare when the EUV lightcurves met the following criteria.

\begin{enumerate}
\item The time of peak \ion{He}{2} emission occurs before the time of peak emission in the log T$_{e}$ = 7.0 temperature bin.
\item The total irradiance of the log T$_{e}$ = 5.8 - 6.2 emission at the time of the 
      \ion{He}{2} peak, must be greater than 3$\sigma$ above the pre-flare average.
\item The derivative of the total log T$_{e}$ = 5.8 - 6.2 irradiance must be negative following the \ion{He}{2} peak.  
\end{enumerate}

\begin{figure}[t]
\centering
\begin{tabular}{cc}
\hspace{-0.2in}\includegraphics[width=0.53\linewidth]{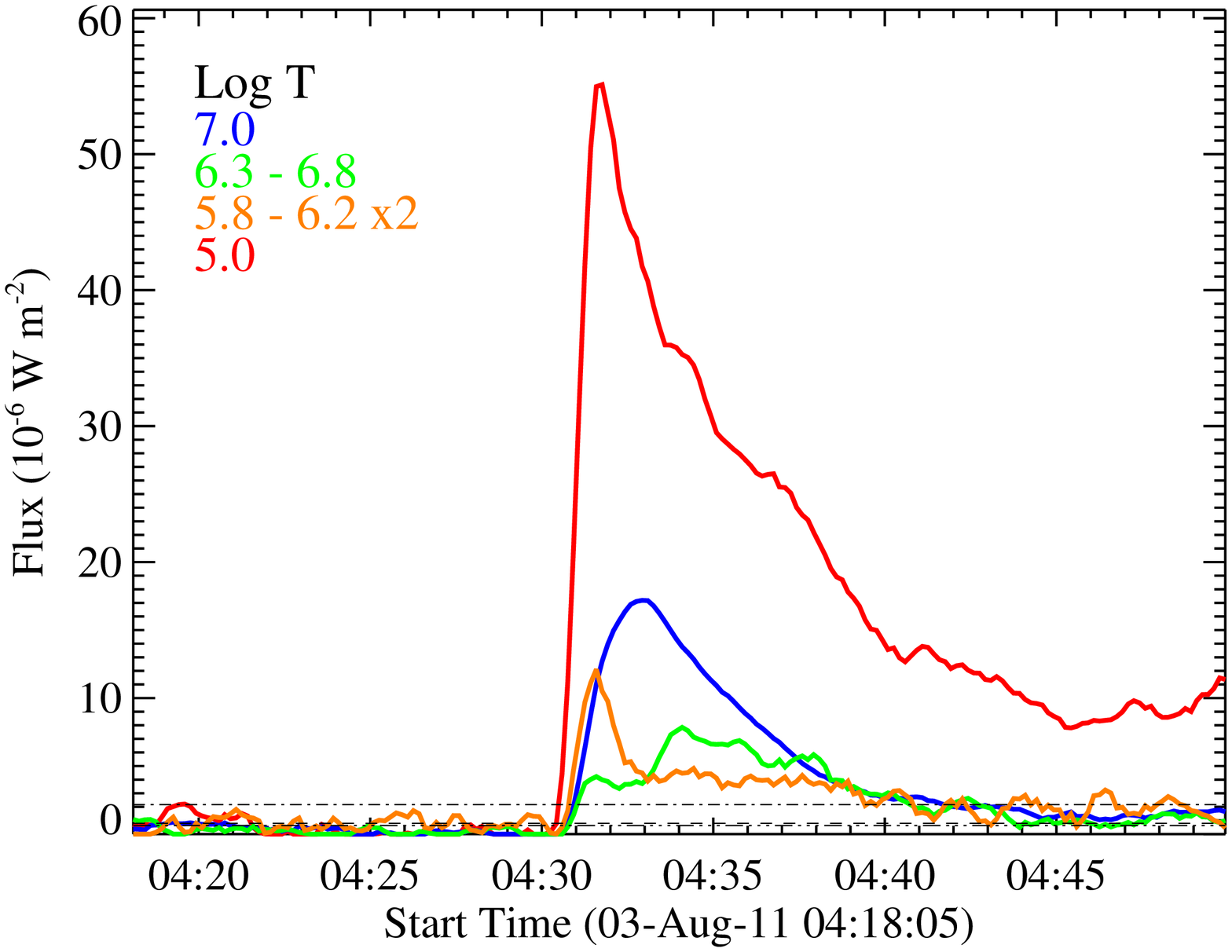} &
 \hspace{-0.3in}\includegraphics[width=0.53\linewidth]{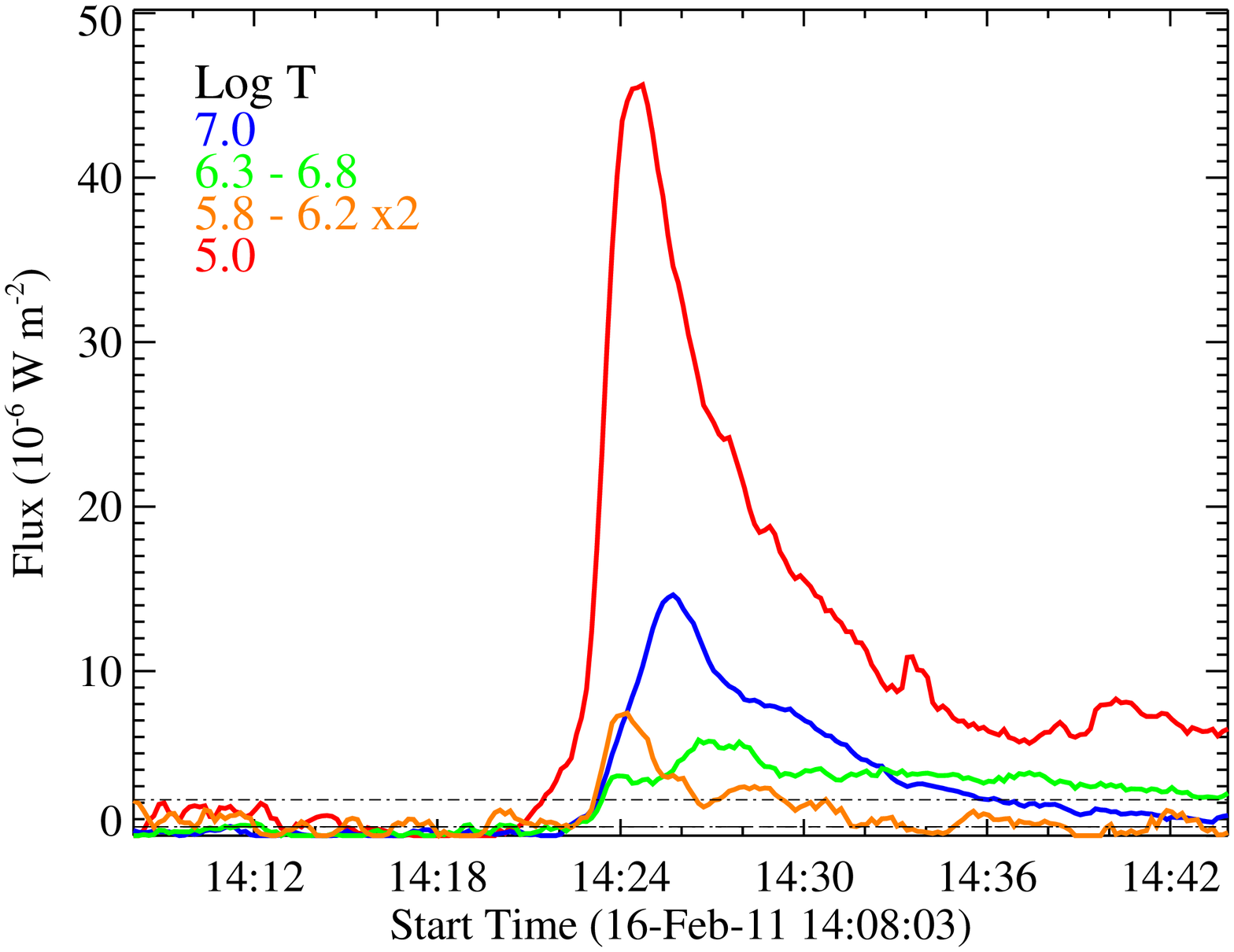} \\
\hspace{-0.2in}\includegraphics[width=0.53\linewidth]{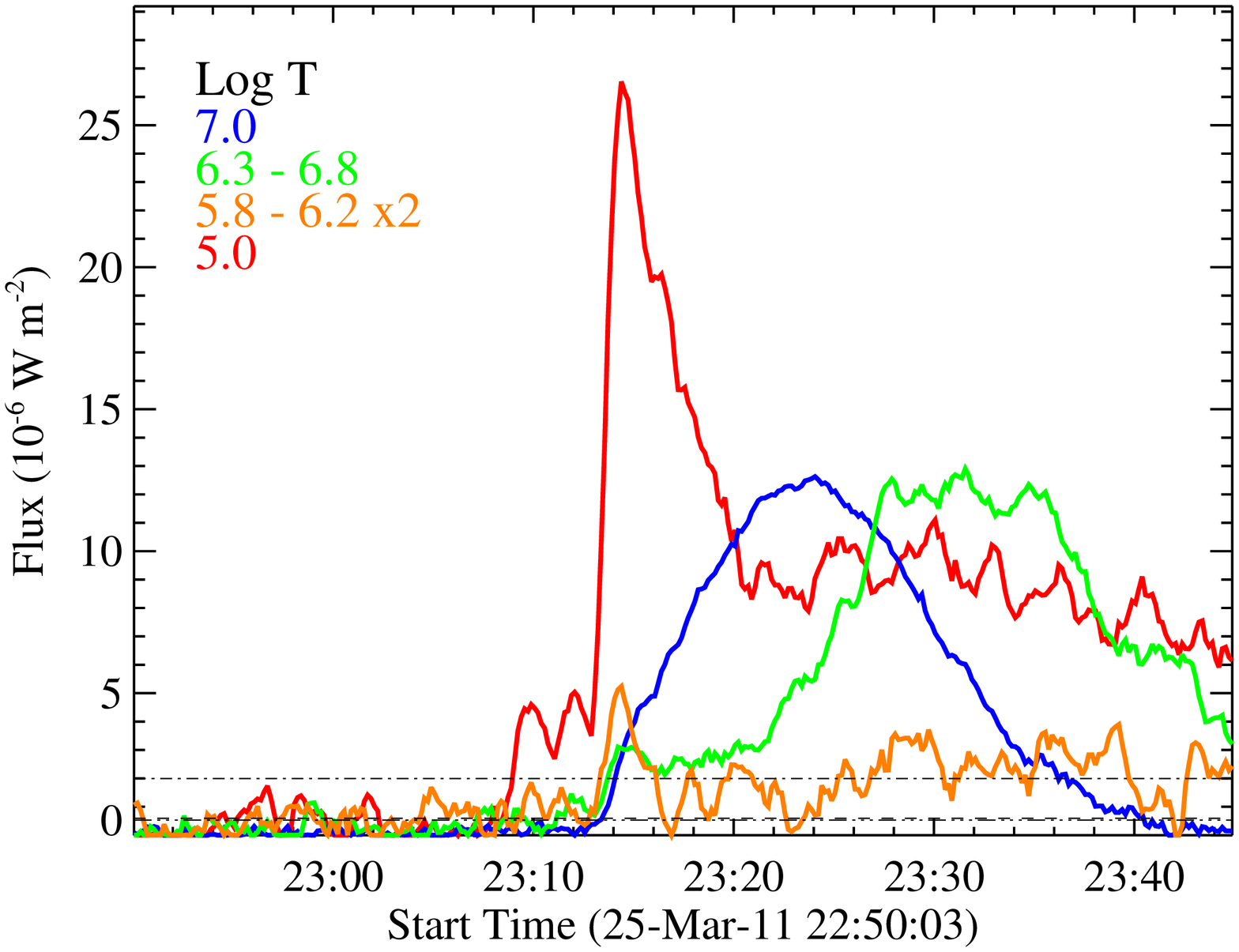} & 
\hspace{-0.3in}\includegraphics[width=0.53\linewidth]{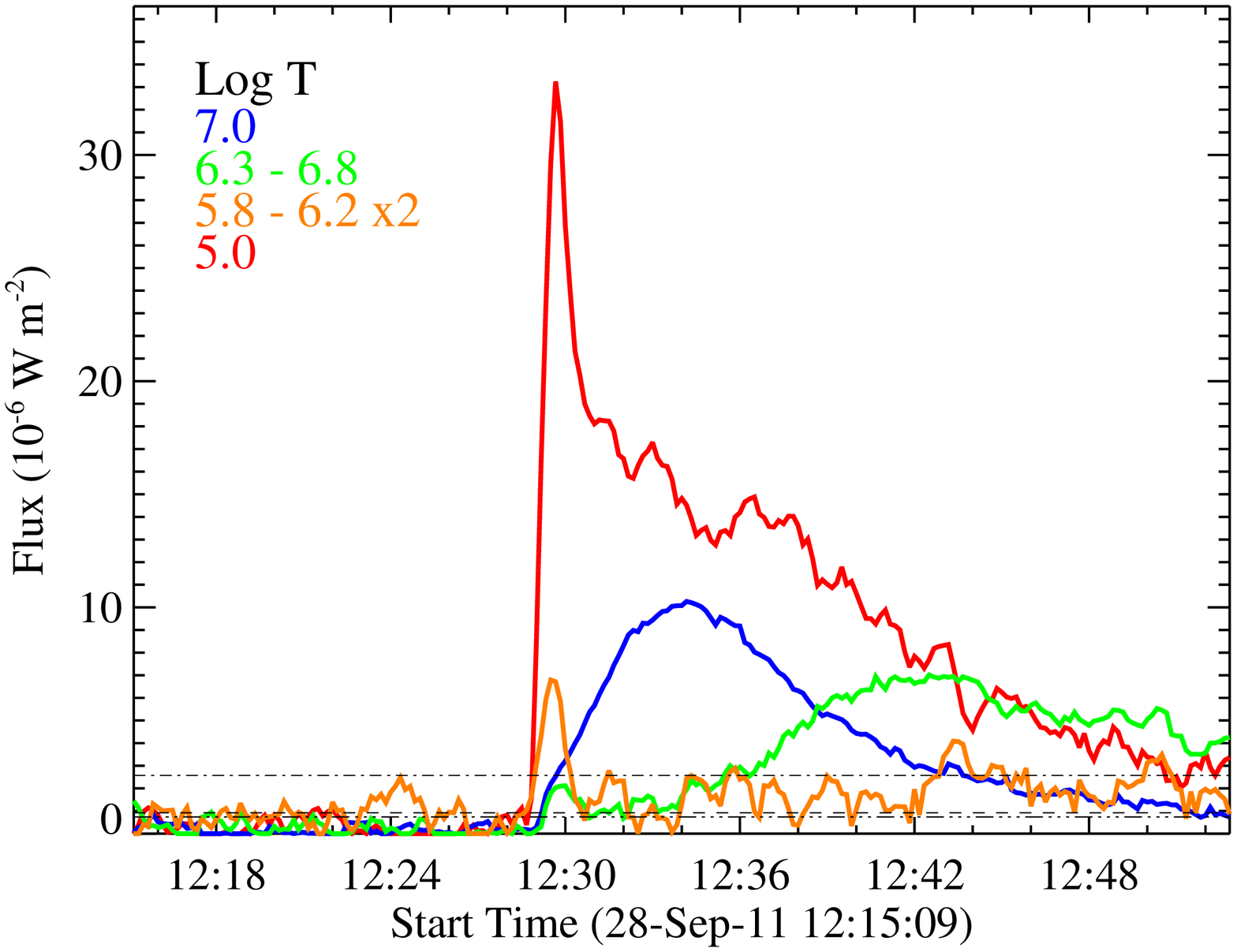} \\
\end{tabular}
\caption{Four examples of flare lightcurves for events which exhibit strong impulsive phase emission. The dashed and dot-dashed lines are the pre-flare average and 3$\sigma$ thresholds for the log T$_{e}$ = 5.8 - 6.2 emission, which have been scaled upwards by a factor of 2 for better visibility.}
\label{f:lc_detect}
\end{figure}

\begin{figure*}
\centering
\begin{tabular}{ccc}
\hspace{-0.2in}\includegraphics[width=0.35\linewidth]{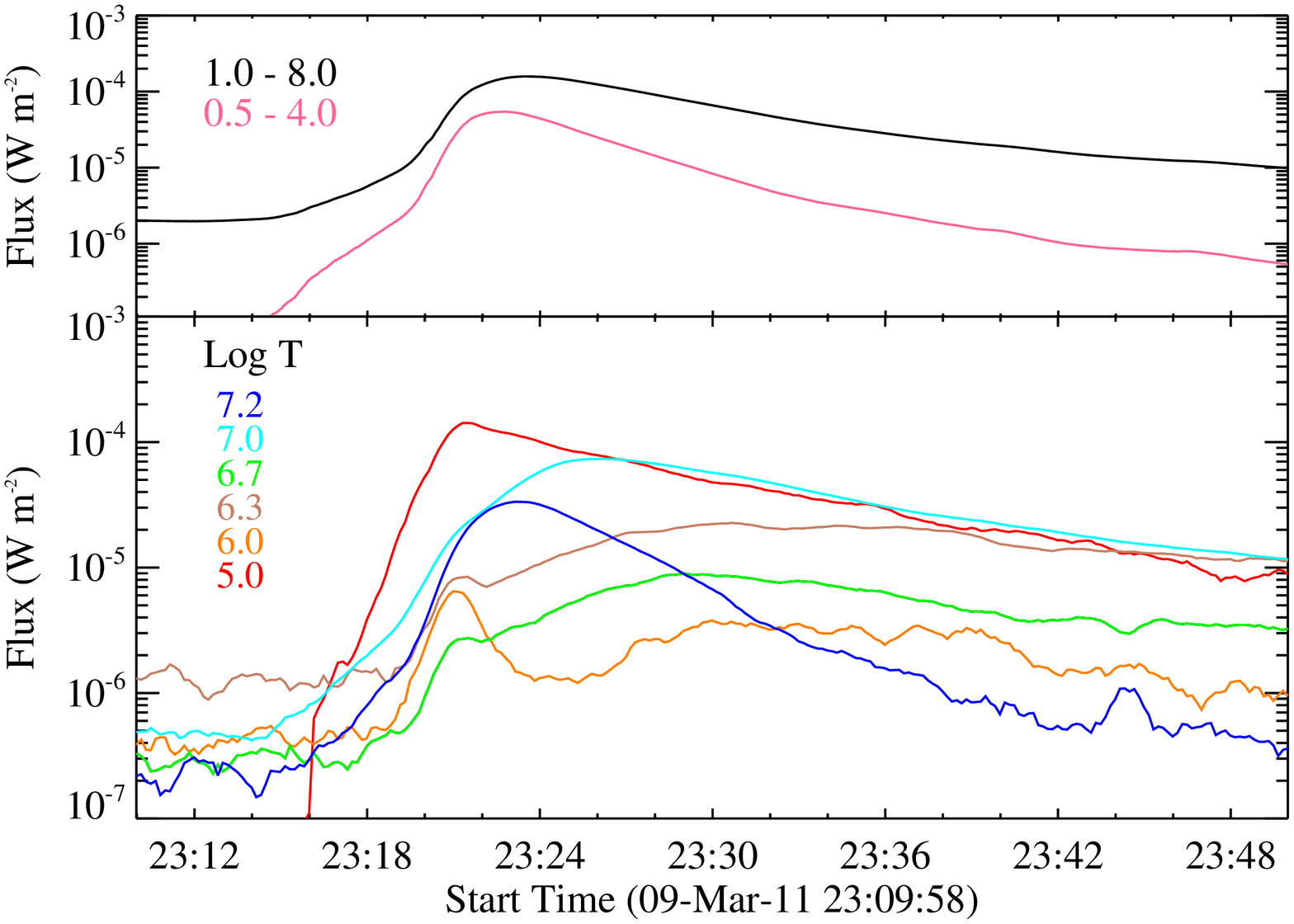}   & 
\hspace{-0.35in}\includegraphics[width=0.35\linewidth]{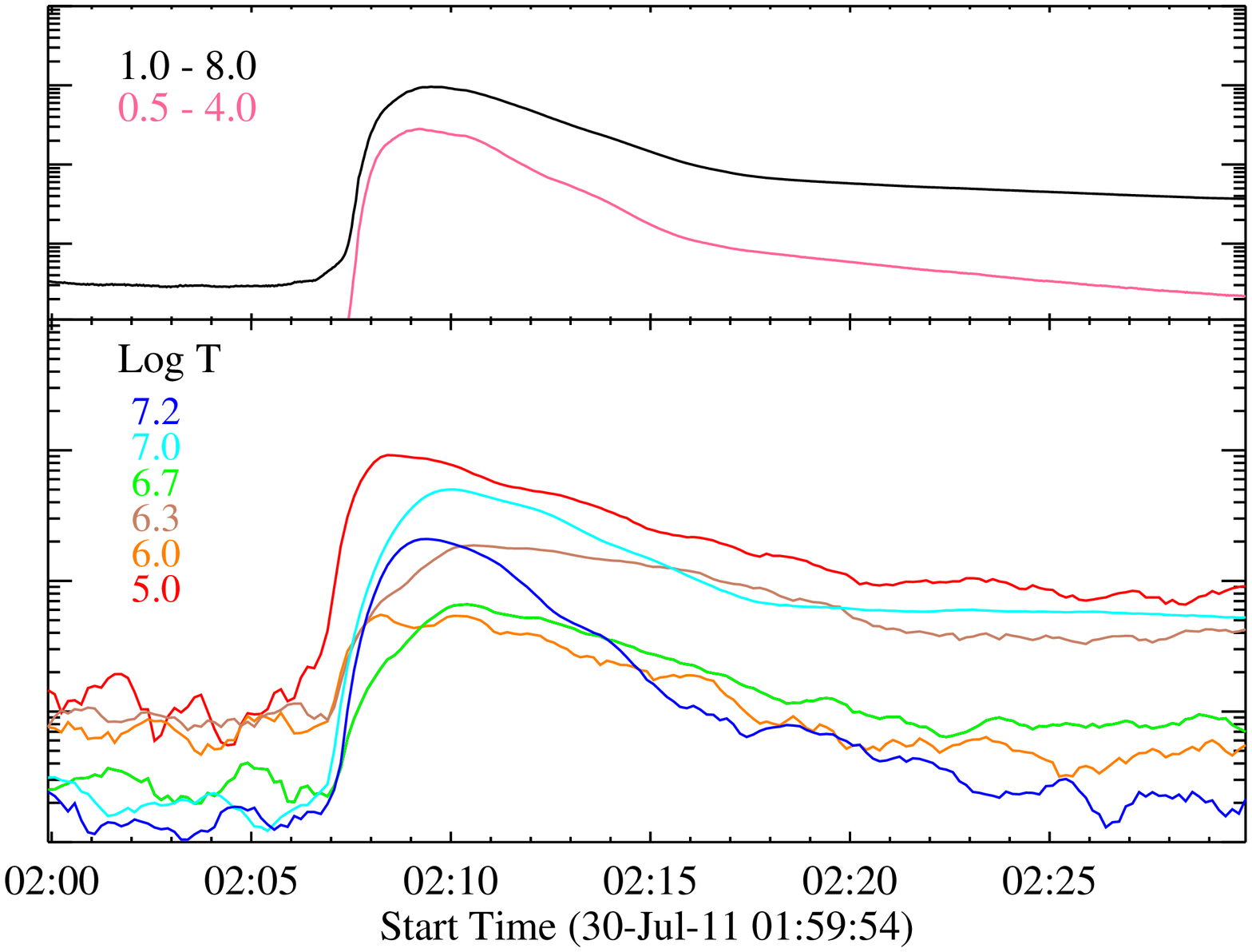} & 
\hspace{-0.35in}\includegraphics[width=0.35\linewidth]{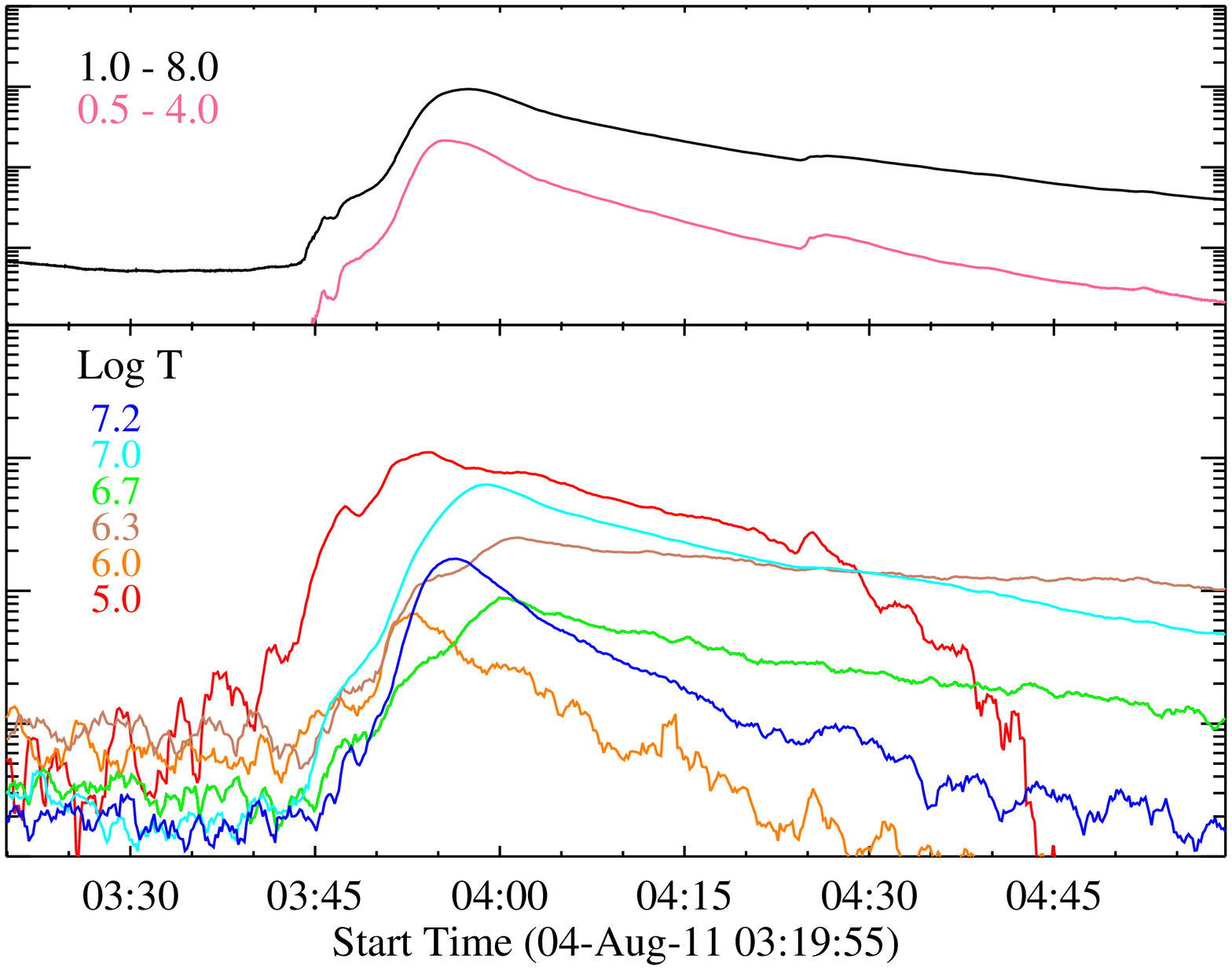}  \\
\hspace{-0.2in}\includegraphics[width=0.35\linewidth]{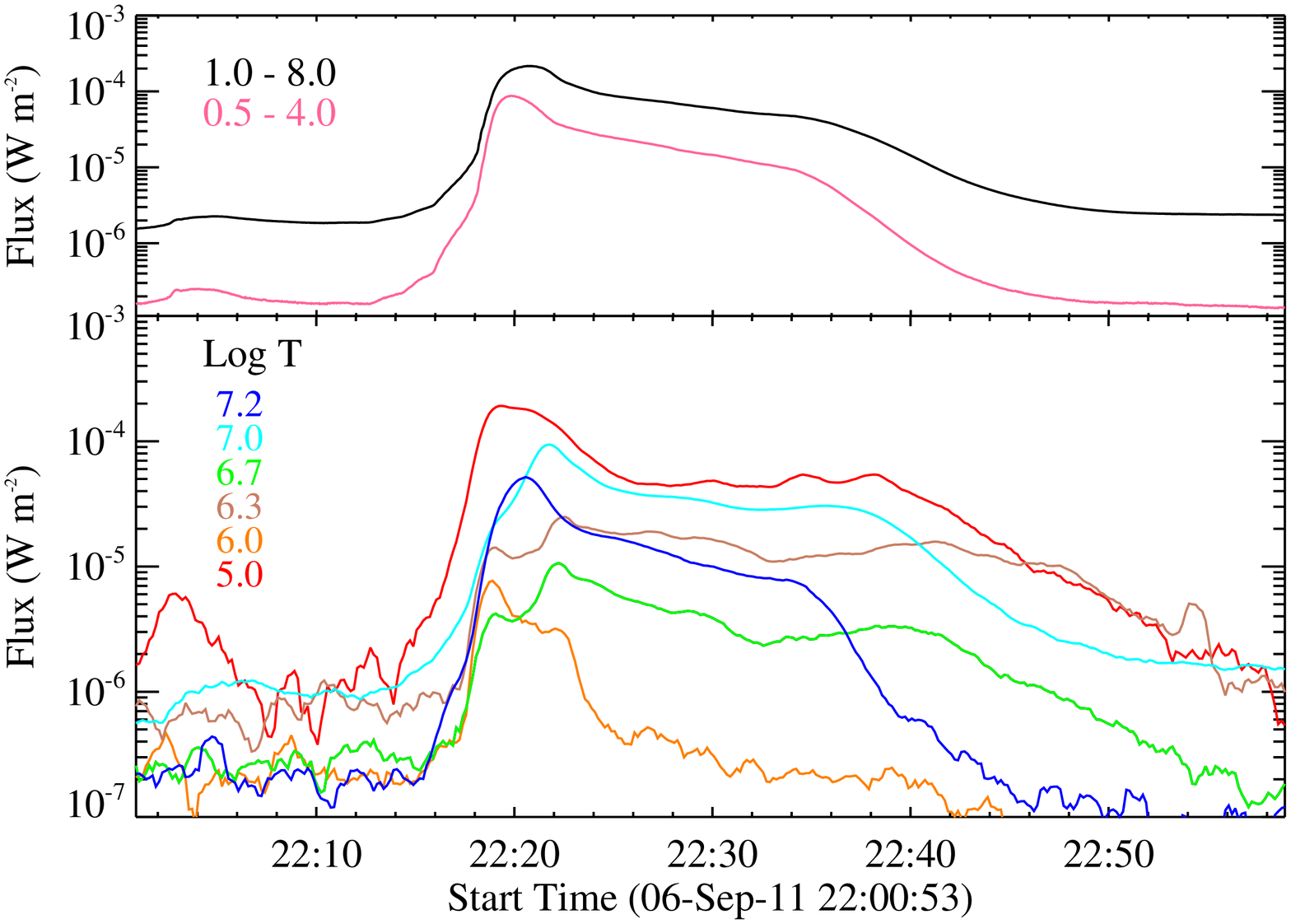}    & 
\hspace{-0.35in}\includegraphics[width=0.35\linewidth]{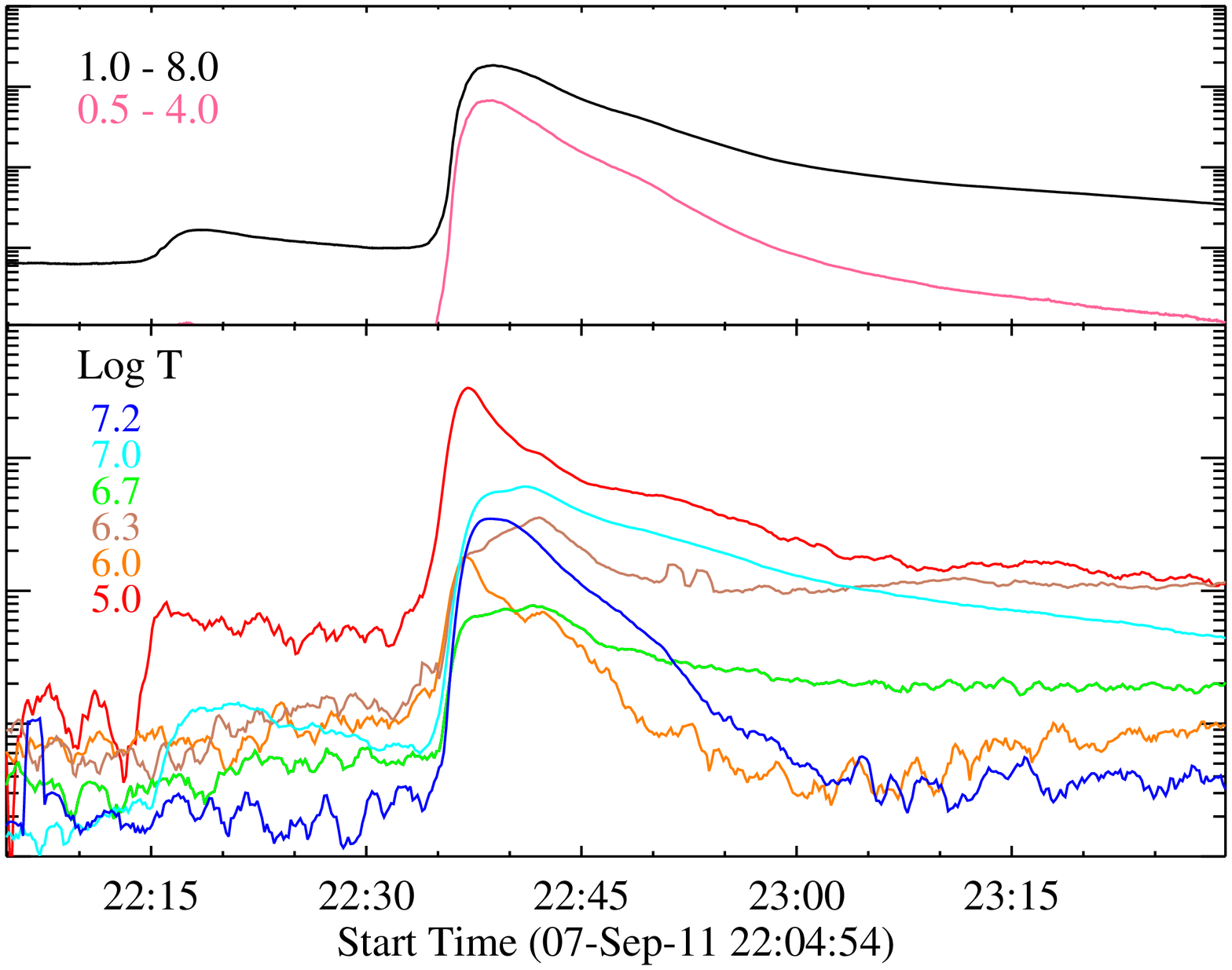}  & 
\hspace{-0.35in}\includegraphics[width=0.35\linewidth]{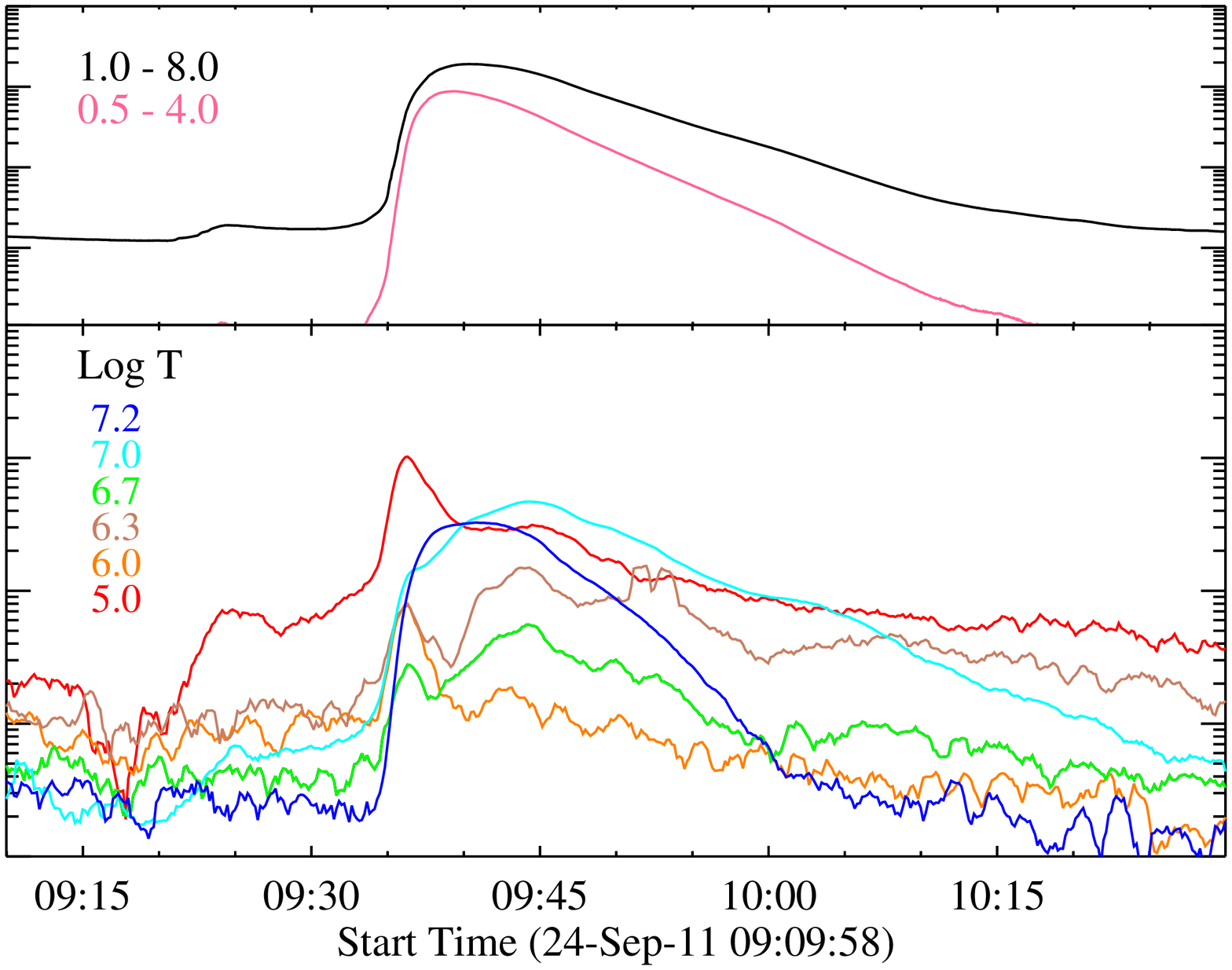} \\
\hspace{-0.2in}\includegraphics[width=0.35\linewidth]{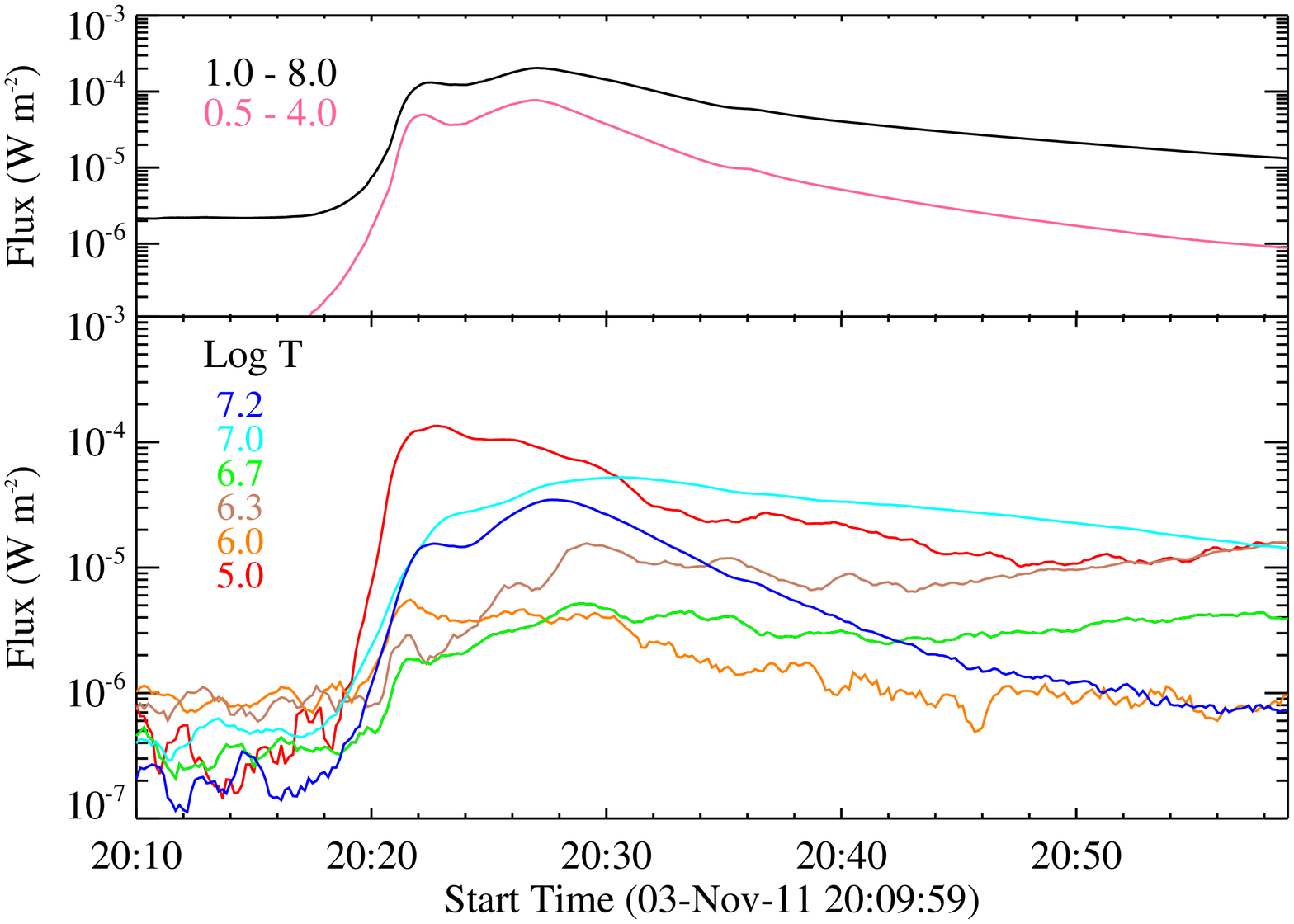} & 
\hspace{-0.35in}\includegraphics[width=0.35\linewidth]{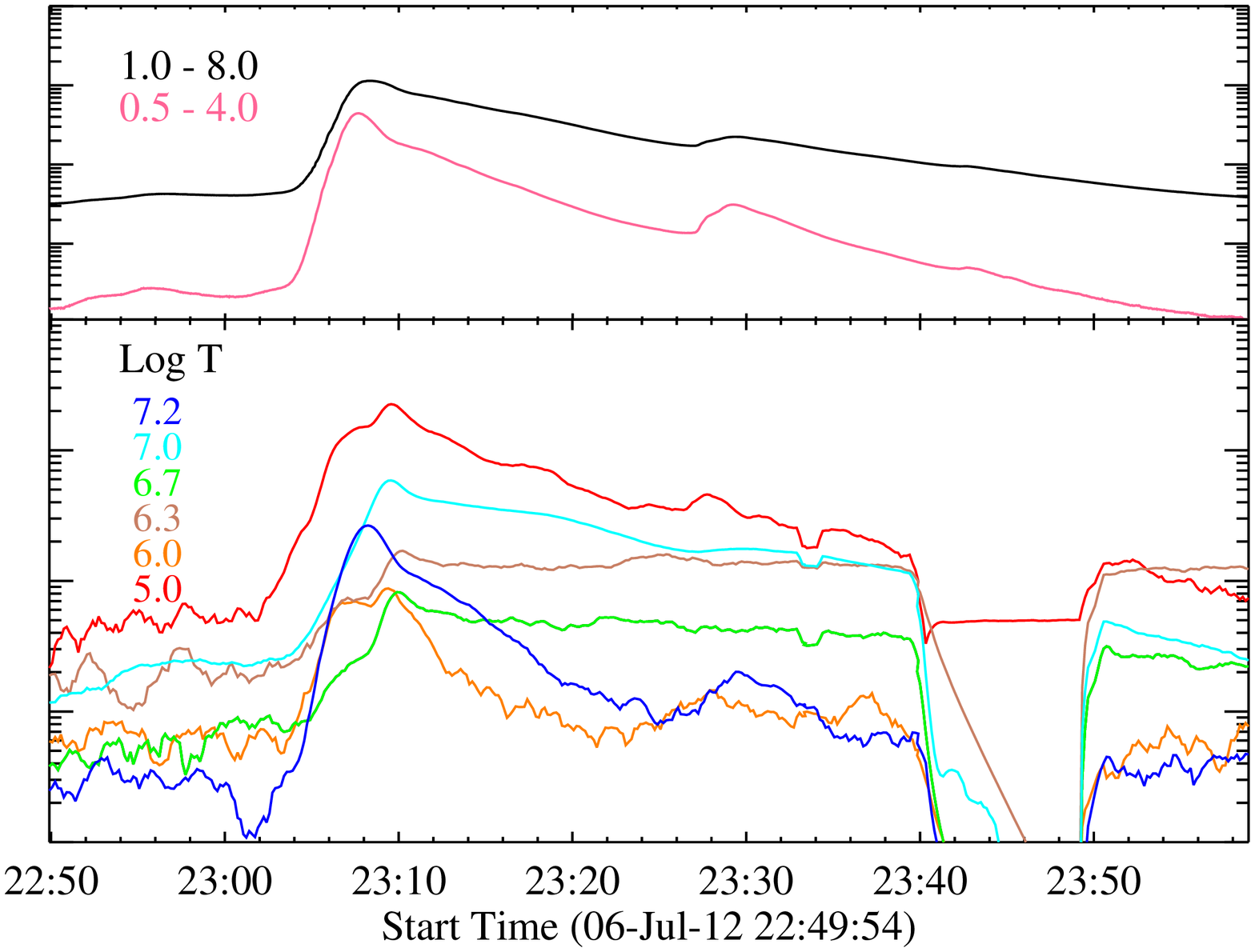} & 
\hspace{-0.35in}\includegraphics[width=0.35\linewidth]{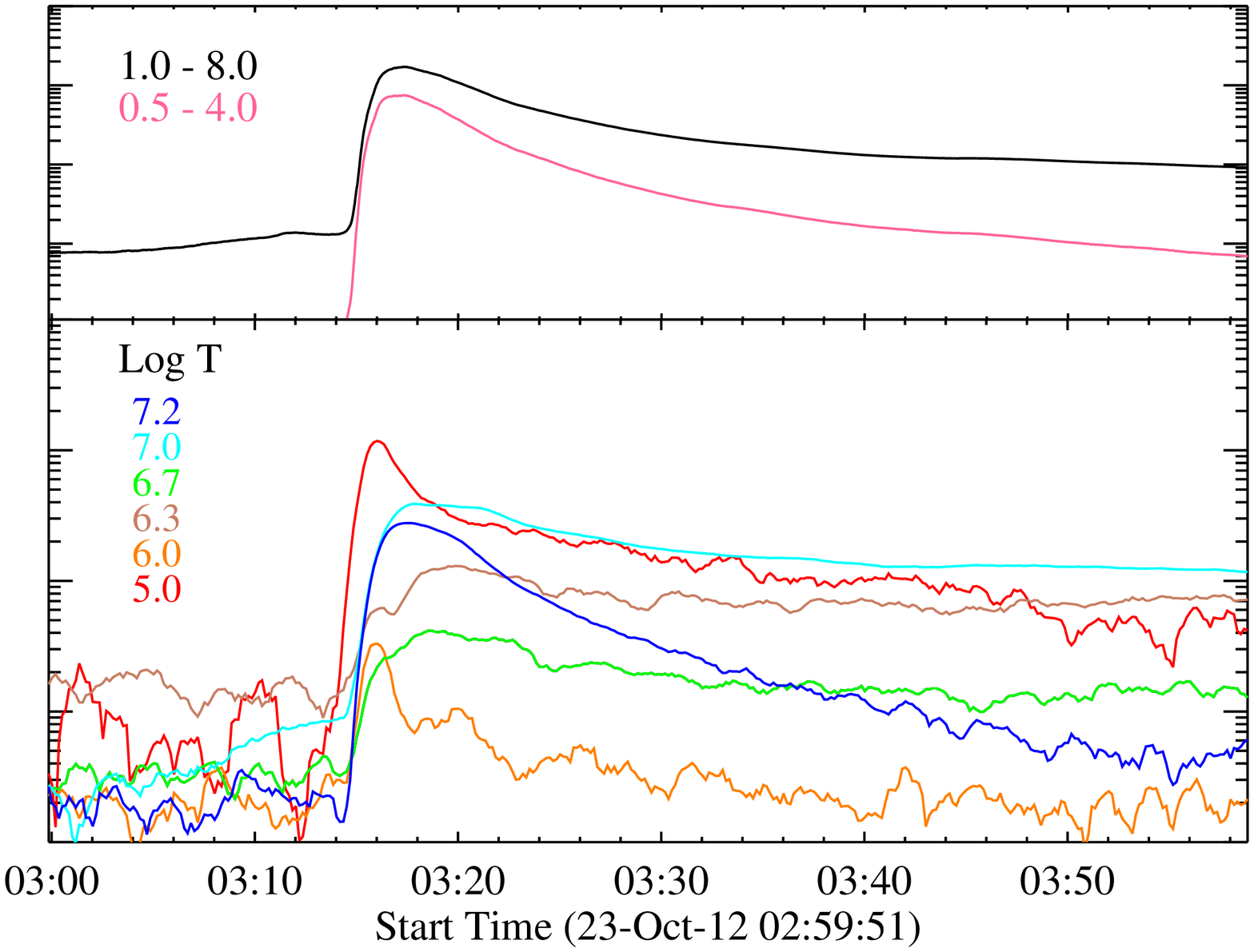} \\
\end{tabular}
\caption{The GOES 1.0 - 8.0\AA\ and 0.5 - 4.0 \AA\ lightcurves are shown for the events studied plotted in black and pink respectively. The colored lines in each bottom panel are lightcurves of emission lines observed by EVE MEGS-A, created by binning the line flux of the emission lines listed in Table \ref{t:linelist} as a function of peak line formation temperature. The log T$_{e}$ = 5.0 temperature bin (red line) is the lightcurve of emission from the \ion{He}{2} 30.4nm doublet. A one-minute boxcar smoothing was applied to the EVE lightcurves for presentation and the $y$-axis range is the same for each plot. }
\label{f:fl_lc}
\end{figure*}

Condition 1 was used to discriminate between `impulsive' flares and long decay events where there may be no easily defined end to the impulsive phase and where the 10 MK emission can be observed to peak before any of the lower temperature emission. Condition 2 rejects flares with no emission that can be observed with EVE above the background, and Condition 3 is used to identify the decline in emission after the end of the impulsive phase, confirming that the increase has an impulsive time profile and was not due to gradual emission or increased pre-flare emission. Four examples of this detection are shown in Figure \ref{f:lc_detect}. From the sample of 390 flares, approximately 100 were found to meet the detection criteria. In this sample the impulsive EUV emission from $\approx$ 0.5 - 3 MK plasma is strong enough that it can be observed above the background emission and can be clearly distinguished from any gradual loop emission or increased pre-flare emission at these temperatures. The highest GOES X- and M-class flares that met the detection criteria were selected for inclusion in the DEM analysis. GOES and EVE lightcurves for these flares are shown in Figure \ref{f:fl_lc}, the latter were created by binning the flux of the emission lines in Table \ref{t:linelist} as a function of temperature. A one-minute boxcar smoothing was applied to the EVE lightcurves for presentation. 

\section{DEM Construction}\label{s:dem}

EVE observations were used to construct the volume DEM (Equation \ref{eq:vdem}) using the Markov-Chain Monte Carlo procedure \citep{kash98} included in the PINTofALE spectral analysis package. The contribution functions for each line were calculated using v7.1 of the CHIANTI atomic database \citep{dere97, land13} with the CHIANTI ionisation equilibrium file, the standard coronal abundances \citep{feld92}, and assuming an electron density of 10$^{11}$ cm$^{-3}$. 

\begin{equation}
DEM(T_{e}) = n_{e}^{2} \frac{\mathrm{d}V}{\mathrm{d} log T_{e}}  \,  
\label{eq:vdem}
\end{equation}

\subsection{Line Selection}\label{ss:lines}

Due to the low spectral resolution of EVE, blending of the many emission lines present at EUV wavelengths is a significant factor that had to be considered for each line used in the analysis. The number of emission lines to be fit changes as the flare evolves and the contributions to an observed emission feature from lines from different elements and ionisation stages will change during a flare.  It is also known that there are inaccurate or missing atomic data for the wavelength ranges observed by EVE and AIA, (e.g. \citet{test12,schm13}). The flare lines in the EVE spectral range have also been the subject of a study by \citet{zann13} who identified blends and lines suitable for EM analyses. The emission lines selected should be the dominant contribution to an observed feature, but there is also uncertainty in measured line irradiance due to the adopted continuum and quiet Sun background. The full line list used is shown in Table \ref{t:linelist} along with the theoretical central wavelength and peak temperature of the line contribution function. However not all of the lines are suitable over the full flare duration. 

There are many emission features in the 9-15nm wavelength range from lines of \ion{Fe}{18} to \ion{Fe}{23} as well as the \ion{Fe}{24} 19.2nm and 25.5nm transitions that can be used to constrain the DEM at high temperature (log T$_{e}$ = 6.9 - 7.2). The \ion{Fe}{23} 13.3nm line has a known blend with \ion{Fe}{20}, while \ion{Fe}{18} 9.4nm line is also blended with \ion{Fe}{20}. An emission feature close to the \ion{Fe}{24} 19.2nm at 19.3nm is a blend of \ion{Ca}{17} and several \ion{O}{5} lines. During the impulsive phase this feature appears to be dominated by \ion{O}{5} emission, but is due to \ion{Ca}{17} in the  gradual phase. The line with highest peak formation temperature in the EVE wavelength range is \ion{Ni}{26} 16.54nm (log T$_{e}$ = 7.4) but this is weak and it is difficult to accurately measure the line irradiance above the continuum emission.

The impulsive phase EUV emission allow the construction of the DEM to be extended down to temperatures of approximately log T$_{e}$ = 5.7 - 6.1 using lines from \ion{Fe}{8} - \ion{Fe}{10}. Given the wavelength range and spectral resolution of MEGS-A it is not possible to observe unblended lines from ions formed at lower temperatures. The feature at 13.1nm is a blend of two \ion{Fe}{8} transitions for which the summed contribution function of both lines is used. The \ion{Fe}{9} 17.1nm line is unblended and is used along with \ion{Fe}{10} 17.4nm. Coronal dimming \citep{hans74, huds96} affects the lowest ionisation states of iron,  \ion{Fe}{8} - \ion{Fe}{14}, and generally emission lines from these ions are not useable after the end of the impulsive phase for the X- and M-class events studied. Even for flares without coronal dimming, emission from lines formed at `quiet' coronal temperatures does not show any significant increase during the gradual phase of the flare.

The impulsive phase DEMs are constructed using emission lines from \ion{Fe}{8} to \ion{Fe}{10}, and \ion{Fe}{14} to \ion{Fe}{24} and the solution obtained is then independent of abundance. Any variation to the adopted iron abundance will only result in the DEM changing by a scaling factor. There is a gap in the temperature coverage from \ion{Fe}{16} at log T$_{e}$ = 6.4 to \ion{Fe}{18} at log T$_{e}$ = 6.9. Emission lines from ions of Ca and Ni which have peak line formation temperatures in this range are present in EVE spectra, but in the impulsive phase the irradiance increase from these lines is too weak to be used reliably.  In the decay phase emission lines from \ion{Fe}{15} to \ion{Fe}{24}, \ion{Ni}{17}, \ion{Ni}{18}, \ion{Ca}{16}, and \ion{Ca}{17} can be used.

\subsection{Line Fitting}\label{ss:line_fit}

As EVE observes the full disk emission it is necessary to perform a quiet Sun background subtraction to obtain the flare emission.  This is an important consideration in the analysis to ensure accurate measurement of the flare excess irradiance, in particular for the lower temperature lines during the impulsive phase for which the irradiance increase is weak. For each event analysed a suitable time range to adopt as the background was initially chosen based on the lowest flux in the GOES 1-8 \AA\ channel in the hour prior to the flare start time. EVE irradiance lightcurves were then checked to ensure that the adopted background interval did not contain any excess EUV emission due to the gradual or late phase emission of a prior flare. The observed irradiance in each wavelength bin was averaged over a 5 minute time interval and subtracted to obtain the excess flare emission. 

Emission lines were fit with Gaussian profiles using non-linear least squares regression. Line fits were weighted by the uncertainty estimate provided for each spectral bin in the Version 3 data release. To account for the variation in free-free emission during solar flares \citep{mill12} and the pseudo-continuum of the many weak lines present, the line flux was measured above the local continuum using a linear background fit over a wavelength range of at least 1nm. The integrated line flux was estimated as the area under the Gaussian profile above the adopted continuum, and uncertainties in line flux were estimated from the statistical uncertainties on the Gaussian profile parameters returned by the least squares fitting. 

\subsection{MCMC Routine}\label{ss:mcmc}

The initial conditions of the PINTofALE Markov Chain Monte Carlo procedure were set to run for 500 simulations, at a temperature binning of 0.05 dex, and the range of the DEM solution allowed to vary over 6 orders of magnitude. The maximum temperature range was log T$_{e}$ = 5.5 - 7.5 when emission lines from \ion{Fe}{8} to \ion{Fe}{24} are used. If lines were excluded, e.g. \ion{Fe}{8} to \ion{Fe}{14} due to coronal dimming, then the DEM construction was performed over a temperature range dependent on the emission lines available to use in the flare spectra at each time bin, and extended to 0.2 dex beyond the lowest and highest peak line formation temperatures. 

\section{Results}\label{s:results}

\begin{figure*}
\begin{center}
\begin{tabular}{ccc}
\vspace{-0.15in}
\hspace{-0.22in}\includegraphics[width=0.33\linewidth]{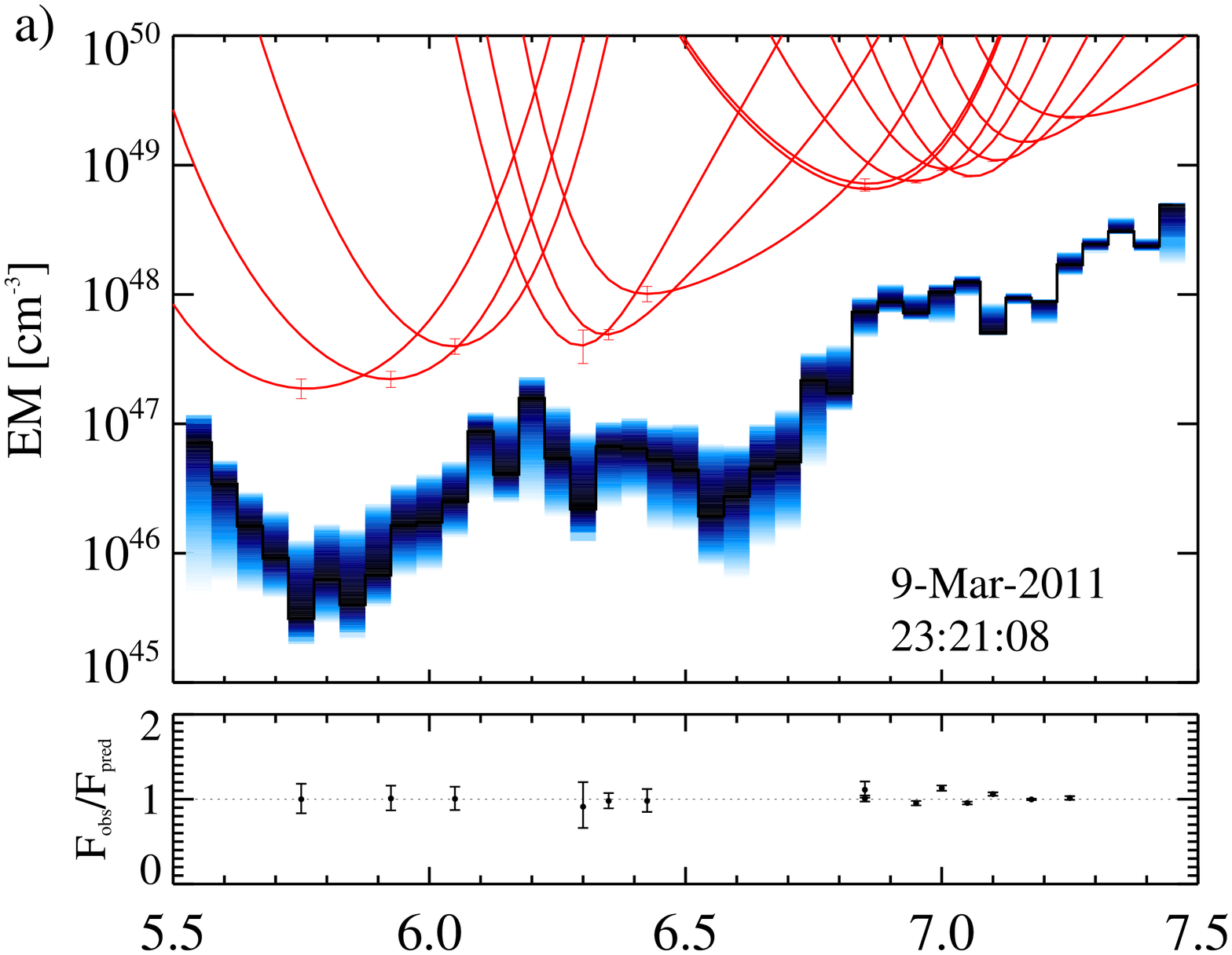} & 
\hspace{-0.22in}\includegraphics[width=0.33\linewidth]{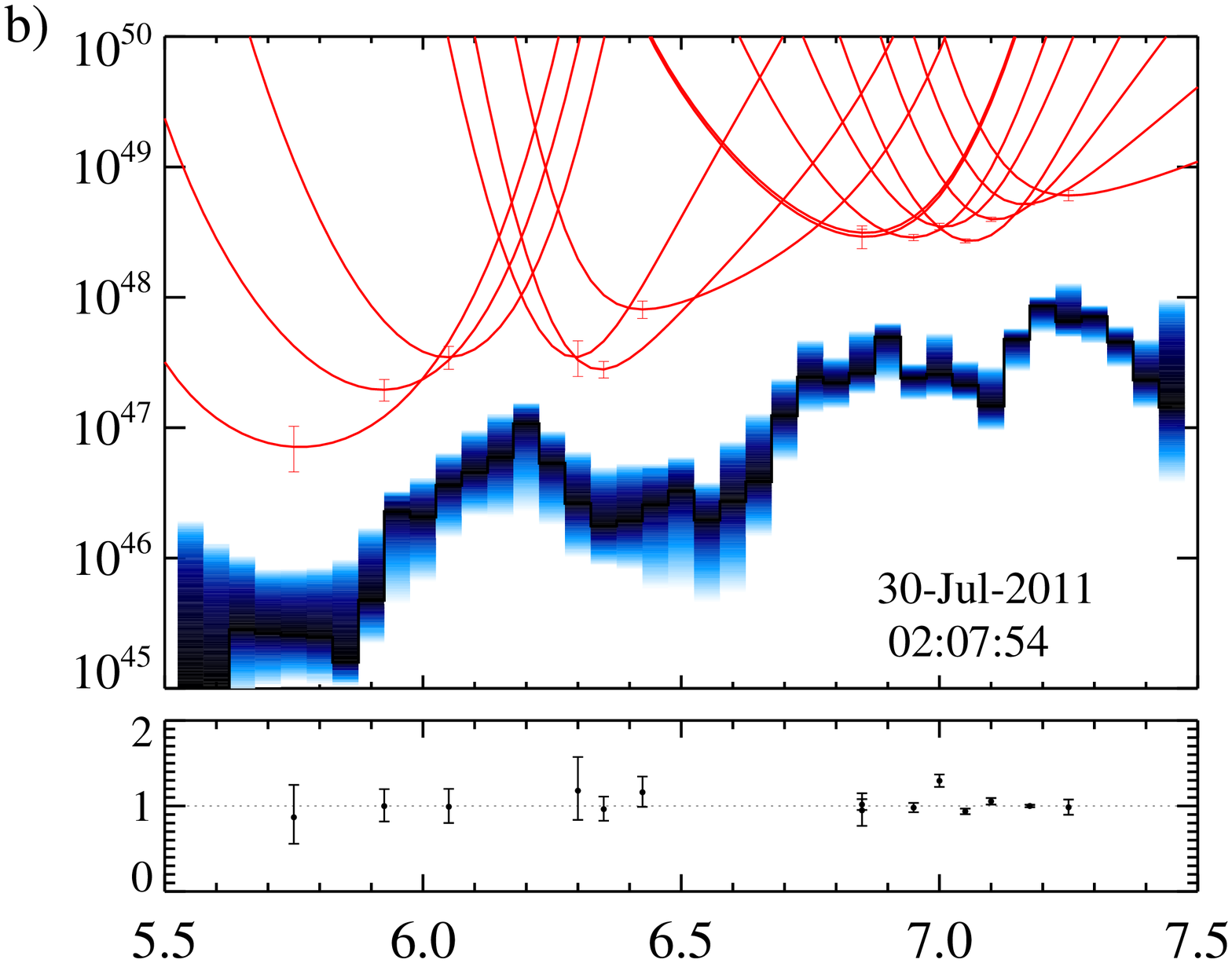} & 
\hspace{-0.22in}\includegraphics[width=0.33\linewidth]{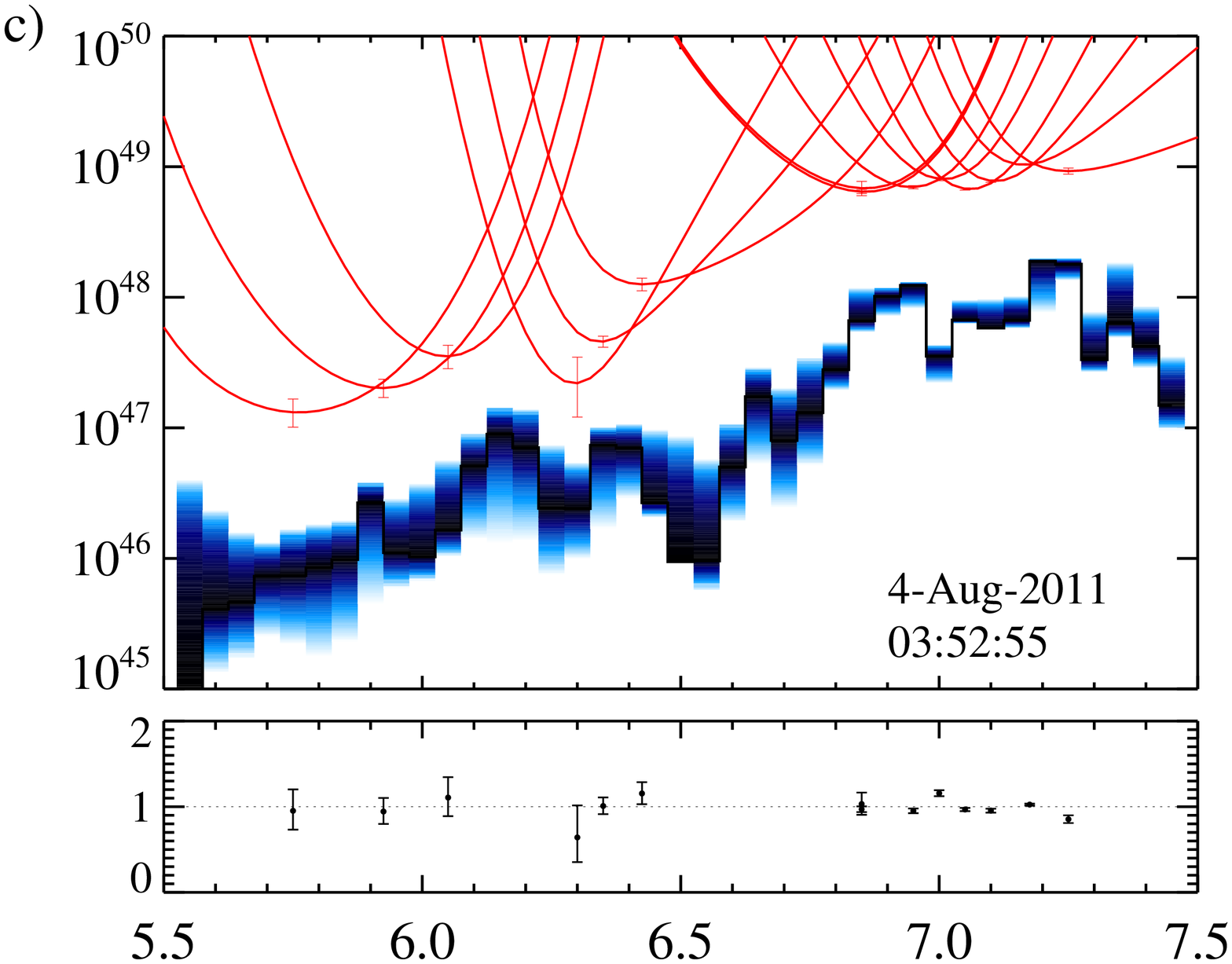} \\
\vspace{-0.15in}
\hspace{-0.22in}\includegraphics[width=0.33\linewidth]{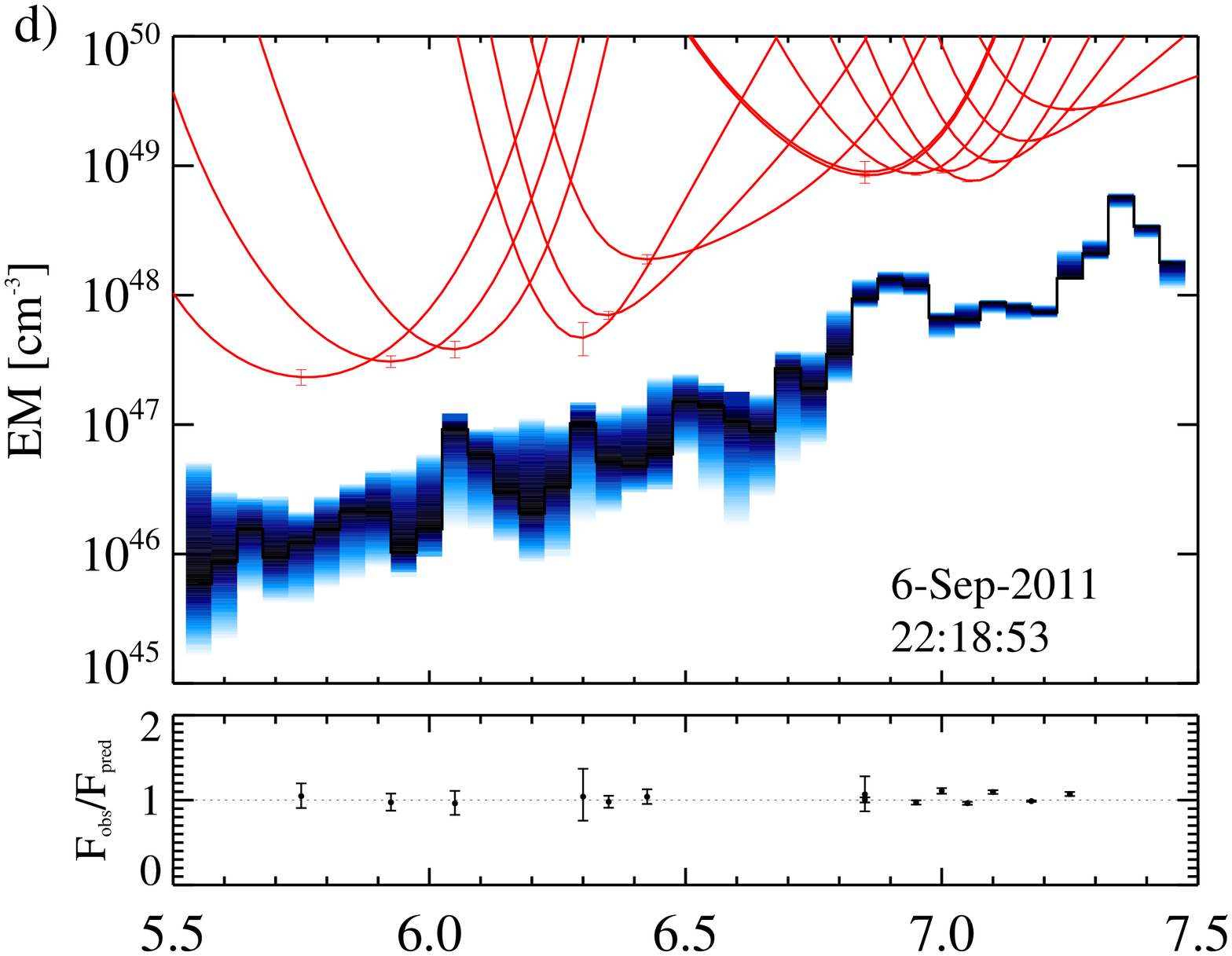} &
\hspace{-0.22in}\includegraphics[width=0.33\linewidth]{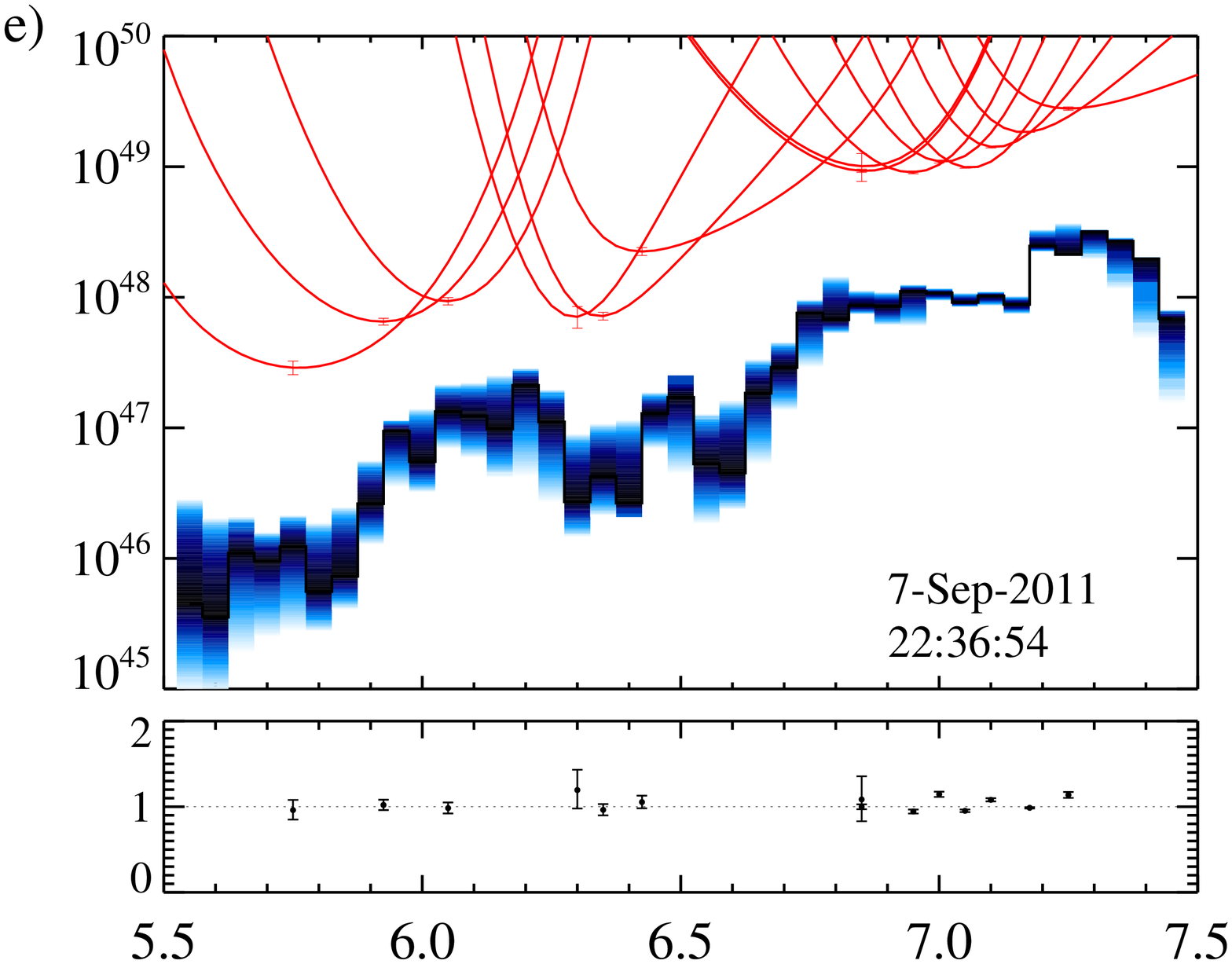}  &
\hspace{-0.22in}\includegraphics[width=0.33\linewidth]{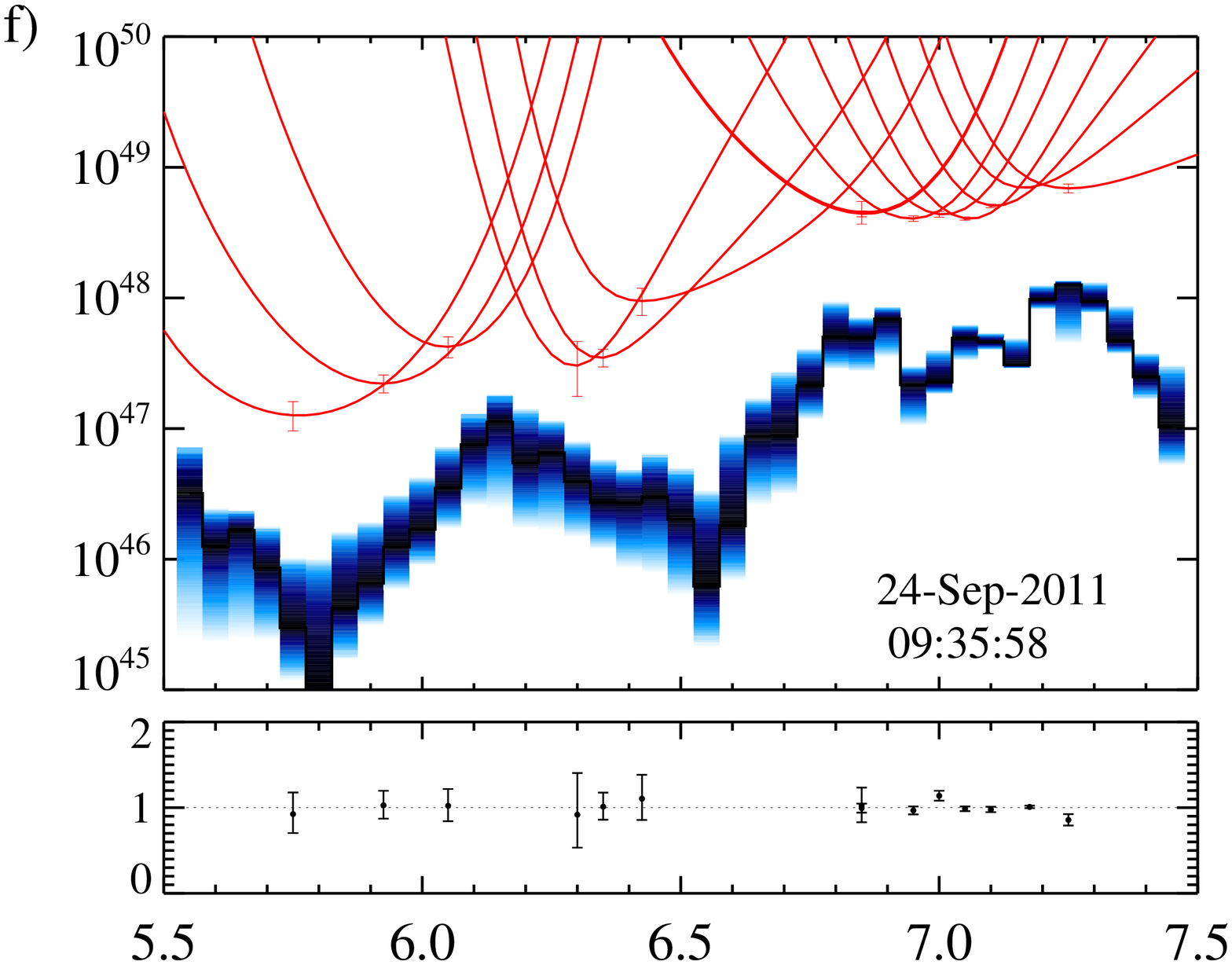} \\
\hspace{-0.22in}\includegraphics[width=0.33\linewidth]{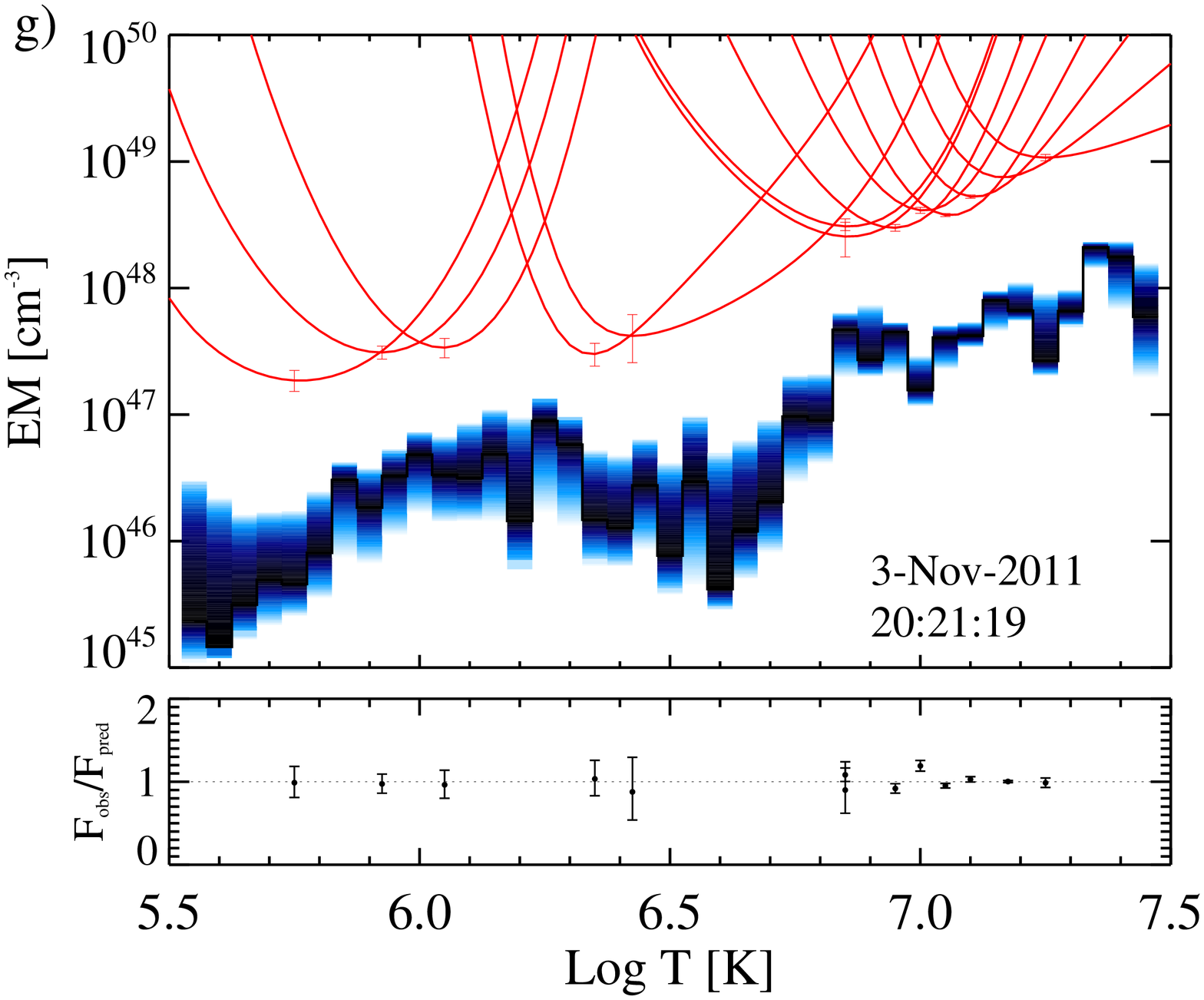} &
\hspace{-0.22in}\includegraphics[width=0.33\linewidth]{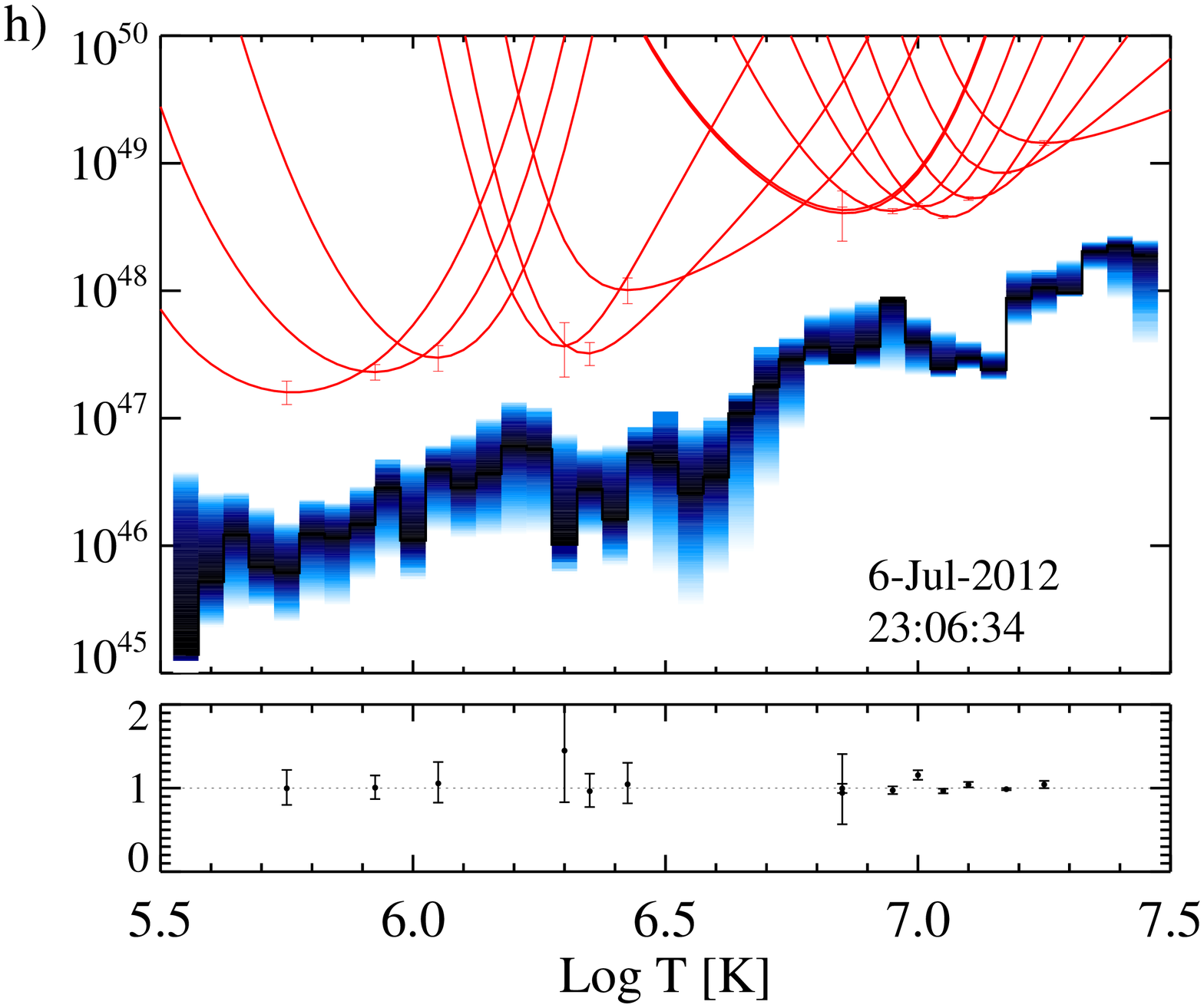} &
\hspace{-0.22in}\includegraphics[width=0.33\linewidth]{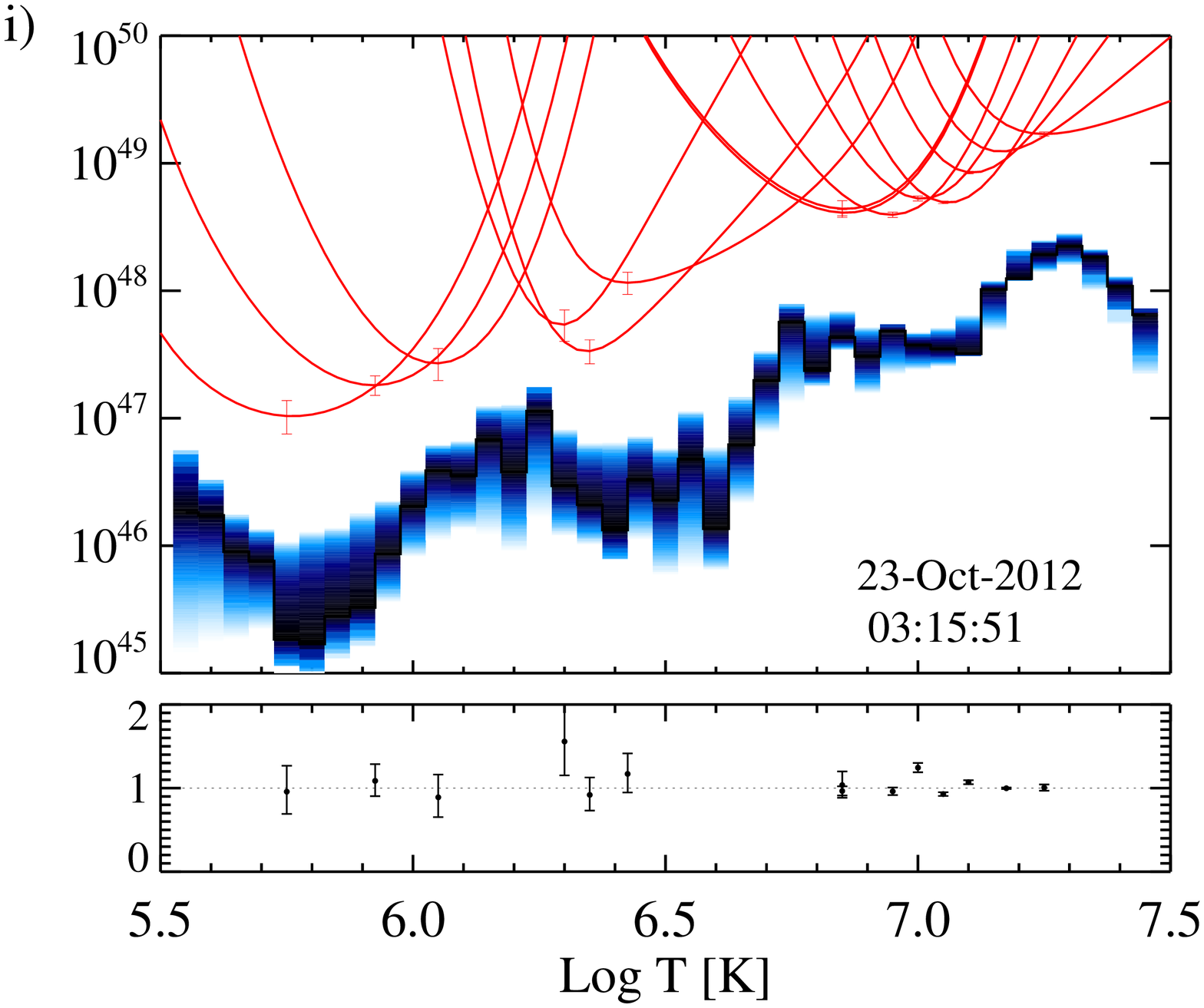} \\
\end{tabular}
\end{center}
\caption{EMDs calculated at the peak of the impulsive phase emission. The best fit solution from the 500 MCMC simulations is plotted as a solid black line with all simulations lying within the 1$\sigma$ confidence bounds plotted shaded in blue. The EM loci curves for each line are over-plotted in red. The ratio of the observed to predicted line flux is shown under each distribution with a dotted line plotted along the one-to-one ratio value. The majority of the flares studied have a two-component structure with a low temperature peak at 1 - 2 MK and a high temperature peak from 7 - 30MK. Several flares (see Panels a), d), e), h) and i)), also exhibit a bimodal distribution peak at high temperature, peaking at approximately log T$_{e}$ = 6.9 and 7.3.}
\label{f:emd_plots}
\end{figure*}

For each of the 9 flares in Table \ref{t:flare_list} the best fit solution from 500 MCMC simulations is plotted in Figure \ref{f:emd_plots} an an emission measure distribution and the EM loci curves showing the maximum possible emission at each temperature are also over-plotted. The ratio of the observed line flux to the line flux predicted by the DEM model is displayed under each EMD.

\subsection{Impulsive Phase}\label{ss:ftpt_dem}

The impulsive phase DEMs are shown for each flare in Fig. \ref{f:emd_plots} which have been constructed using line fluxes at the peak of the impulsive phase emission. For each of the nine events studied a similar distribution is found. At low temperatures the DEM increases from a few 100,000 K and reaches a peak at approximately 1 - 2 MK. The DEM then decreases and reaches a minimum at 3 - 4 MK. After this minimum, the DEM increases again and there is a broad, high temperature component in the solution from 7 - 30 MK. Several events also exhibit a double peaked distribution at high temperature, with peaks at log T$_{e}$ = 6.9 and 7.3 (8 and 20 MK). In Table \ref{t:flare_emt} the peak temperature and total emission measure of the low temperature component for log T$_{e}$ $\le$ 6.6 is shown. The re-constructed DEMs broadly reflect what is seen from the flare EUV lightcurves;  strong, impulsive phase increases are observed from ions up to \ion{Fe}{10}, with weaker increases detected from ions up to \ion{Fe}{16}. There appears to be a deficit of material at temperatures intermediate to quiet Sun coronal temperatures at $\approx$ 1 MK,  and the high temperature flare emission from $\approx$ 10 MK plasma.

\subsection{DEM Verification}\label{ss:dem_test}

The re-constructed DEMs using the MCMC method give good agreement between the supplied and predicted line irradiance. The ratios of the observed to predicted line irradiance are generally in the range 0.8 - 1.2. Each line was also checked to look for systematic errors in the choice of emission lines used in the analysis.  It was found that the measured flux of the \ion{Fe}{20} 12.18nm emission line was greater than that predicted by the DEM model for every event studied. The measured line flux was found to be on average 17$\%$ greater than the predicted flux. Synthetic spectra were generated from the constructed DEMs and the CHIANTI database as an additional check to help identify a possible blend, but do not show any other strong emission lines within the \ion{Fe}{20} line profile. It is possible an unidentified emission line could be contributing to the measured line flux. The observed flux of the \ion{Fe}{14} 21.1nm line also tends to be higher than the model prediction, however this line has large observational uncertainties.

\begin{figure}
\begin{tabular}{cc}
\hspace{-0.04in}\includegraphics[width=0.5\linewidth]{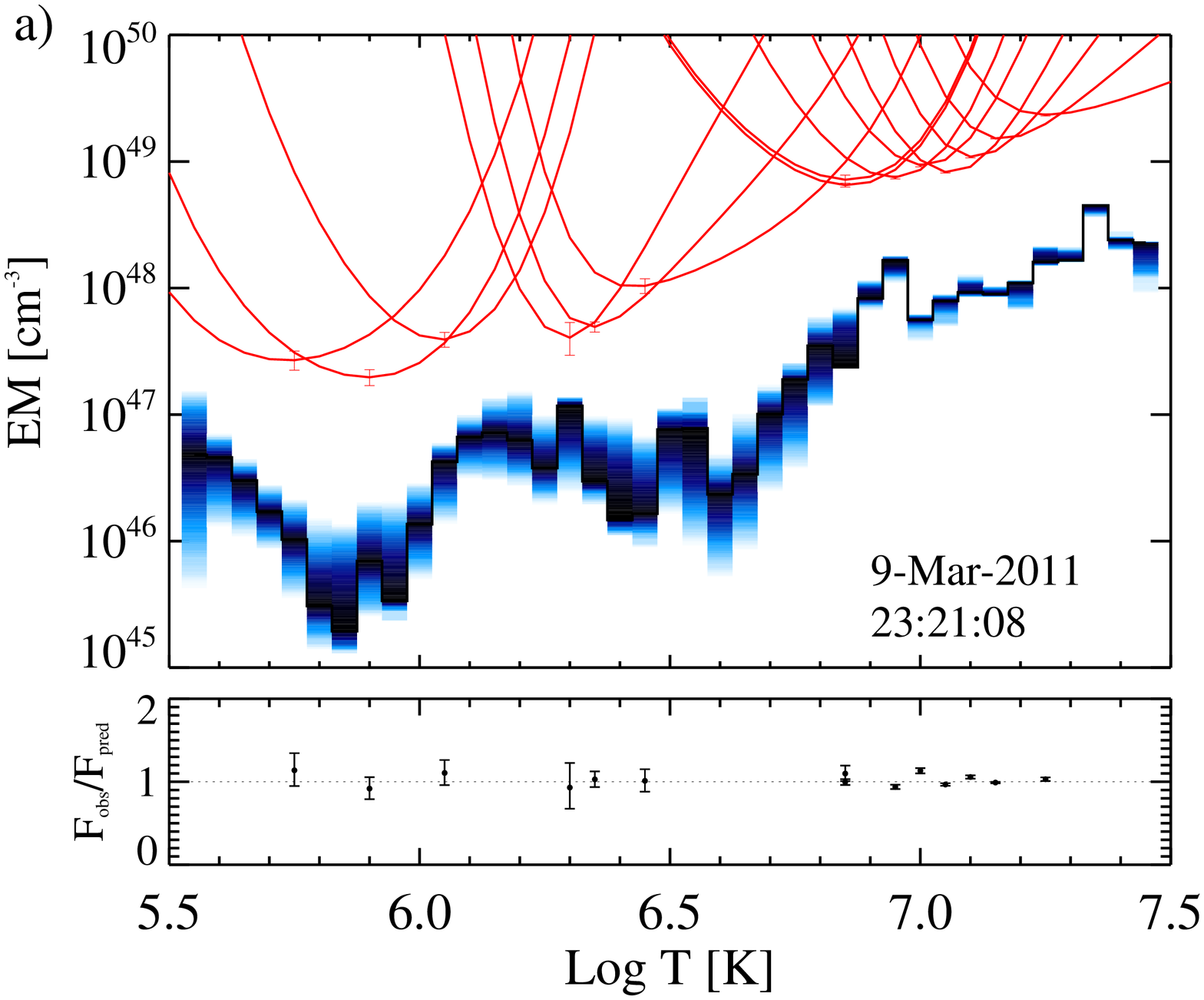} &
\hspace{-0.21in}\includegraphics[width=0.5\linewidth]{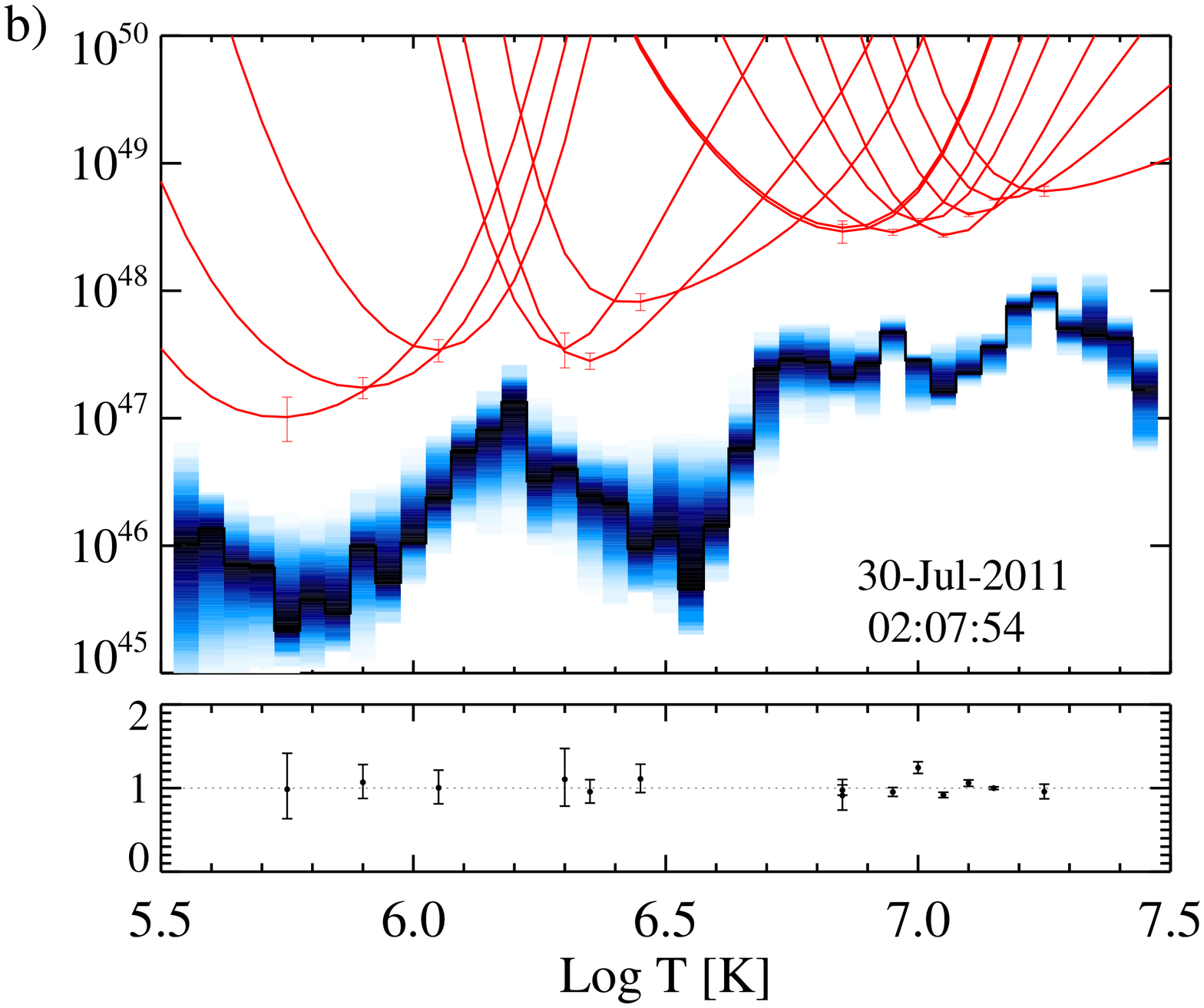} \\
\end{tabular}
\caption{The results of the MCMC method obtained when using the \citet{brya09} ionisation equilibrium data. There is little difference in the overall solution compared to the DEMs obtained when using the CHIANTI data.}
\label{f:dem_bryans}
\end{figure}

To investigate whether the choice of ionisation equilibrium data or adopted DEM method had any affect on the solution obtained, several checks were performed to verify the structure in the re-constructed DEMs. Line contribution functions were also calculated using the \citet{brya09} ionisation equilibrium data, with the best fit solutions for two flares shown in Fig. \ref{f:dem_bryans}. The major difference in the ionisation equilibrium data is the ion fraction for \ion{Fe}{8}, but there are little discrepancies between the two solutions.

A different construction method to obtain the DEM was also investigated. The NRL-EVE method \citep{warr13} was used which assumes a parameterised form of the DEM as a sum of Gaussians. The wavelength ranges used in the least-squares minimisation were altered to cover the \ion{Fe}{9} and \ion{Fe}{10}  emission lines from 17 - 18 nm, and the \ion{Fe}{14} 21.1nm line. The temperature range was set to cover log T$_{e}$ = 5.5 - 7.5 with 10 Gaussian components, and, following the procedure outlined in \citet{warr13}, the underlying continuum was removed by subtracting the lowest irradiance in each 1nm interval from the observed spectra. 

\begin{figure}
\begin{tabular}{cc}
\vspace{-0.18in}
\hspace{-0.04in}\includegraphics[width=0.4\linewidth]{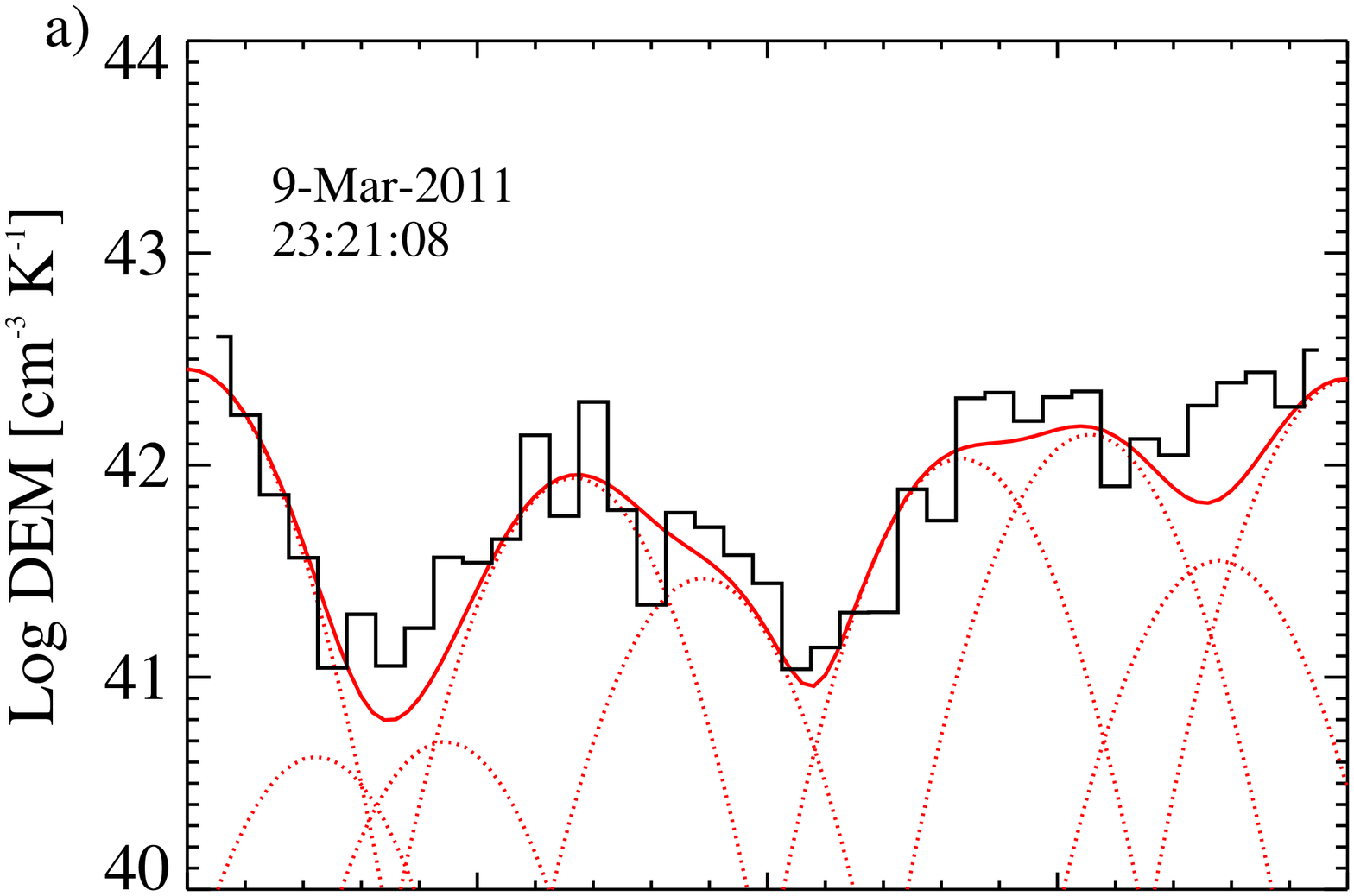} &
\hspace{-0.21in}\includegraphics[width=0.4\linewidth]{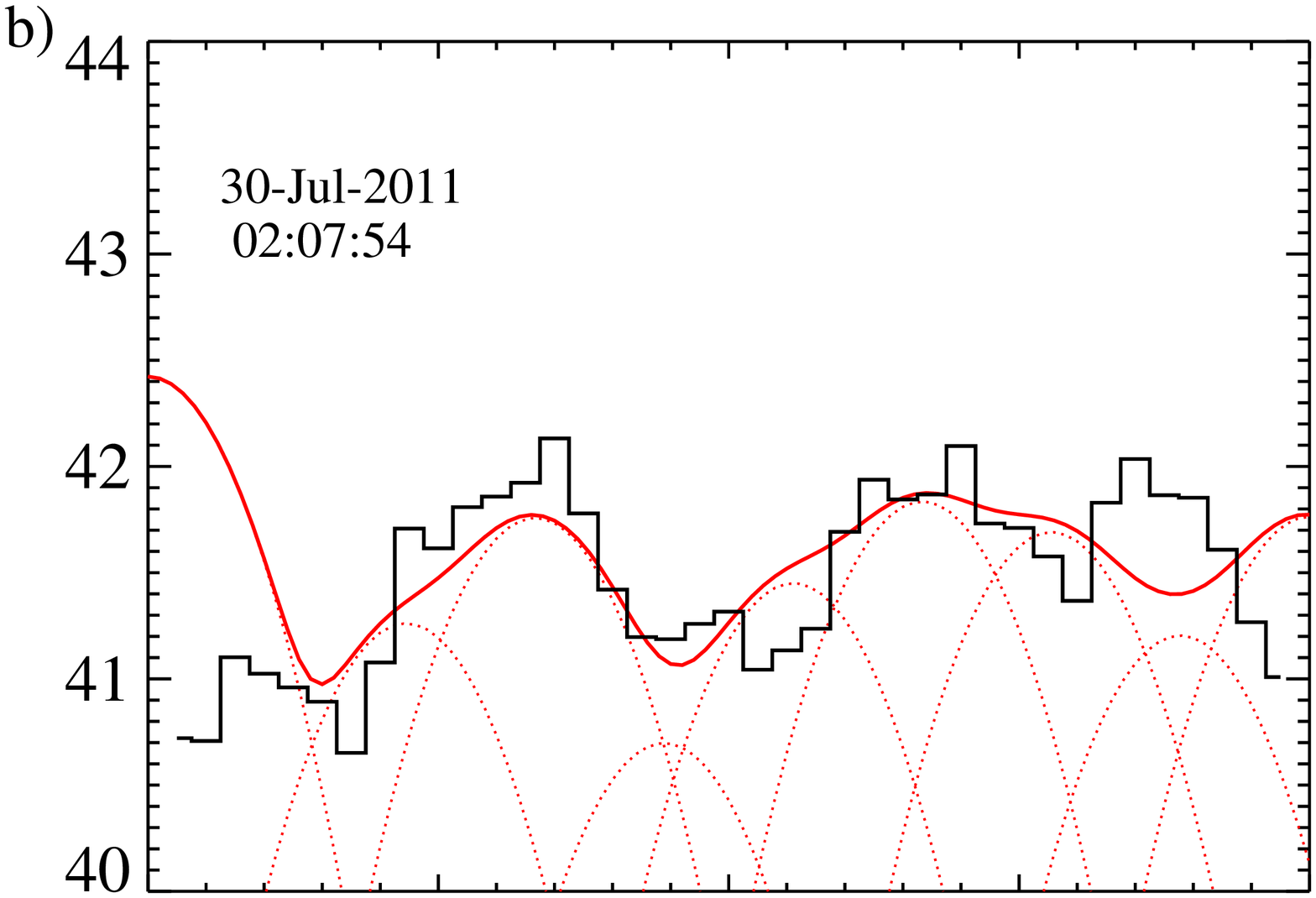} \\
\vspace{-0.18in}
\hspace{-0.04in}\includegraphics[width=0.4\linewidth]{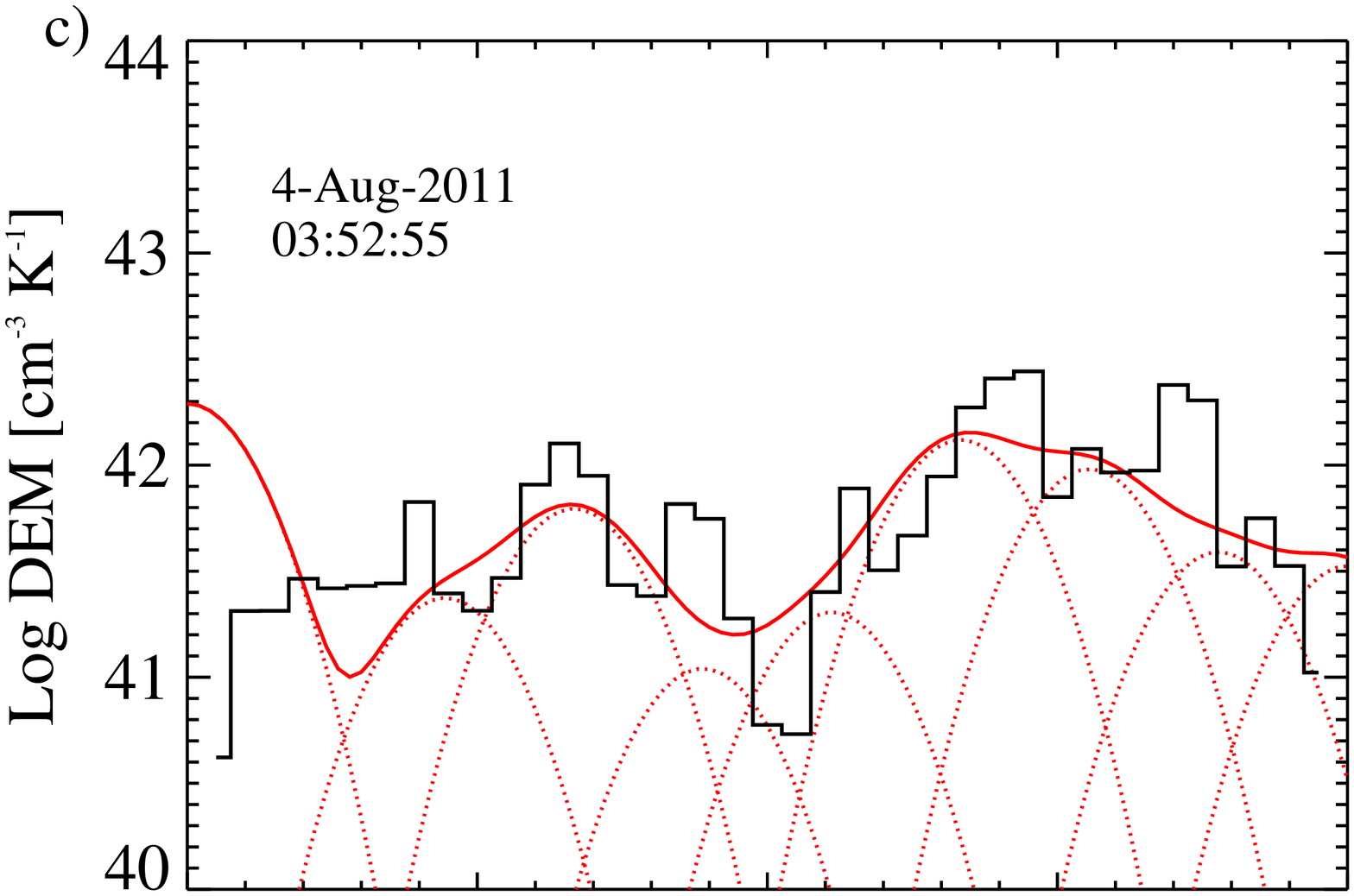} &
\hspace{-0.21in}\includegraphics[width=0.4\linewidth]{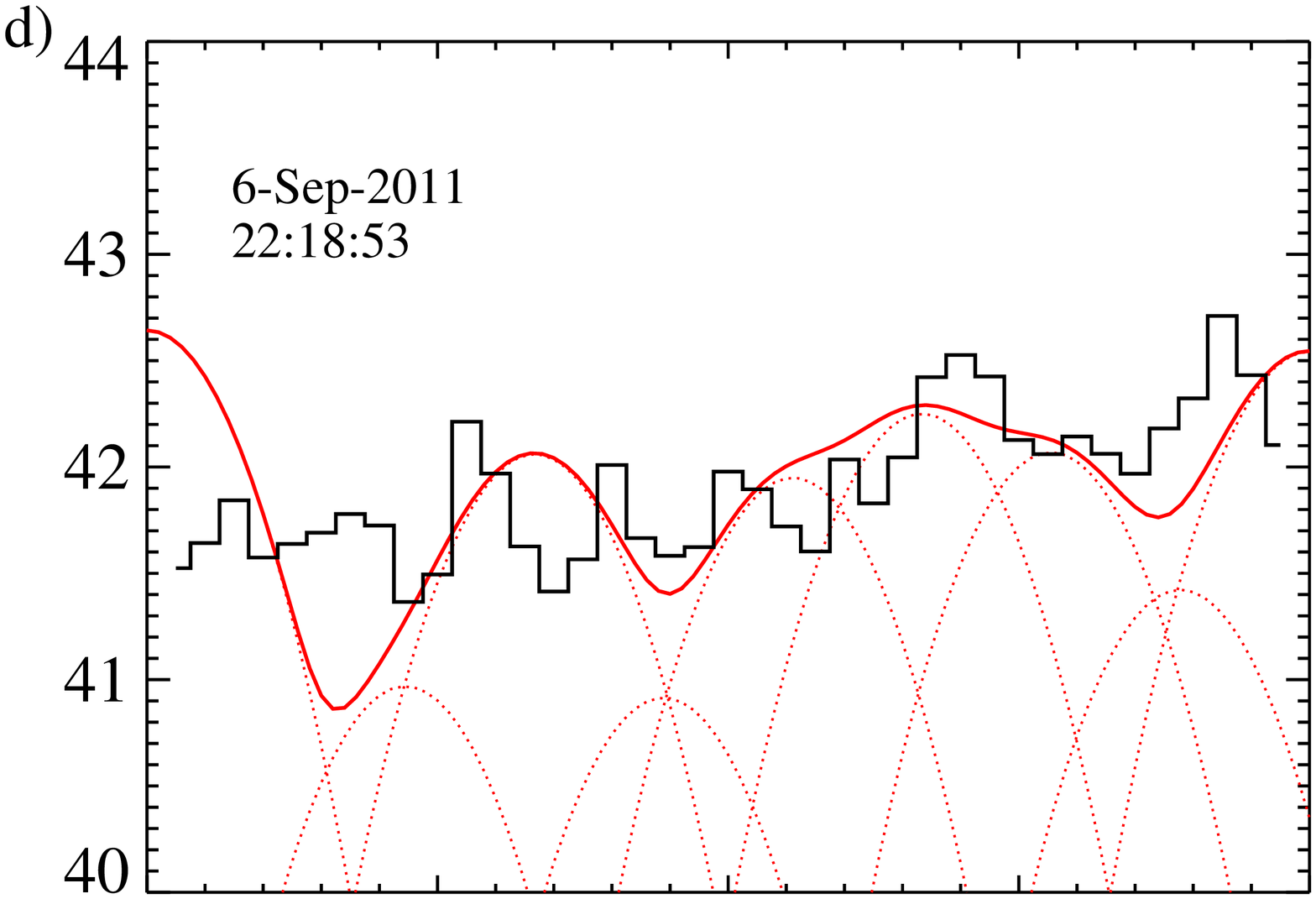} \\
\vspace{-0.18in}
\hspace{-0.04in}\includegraphics[width=0.4\linewidth]{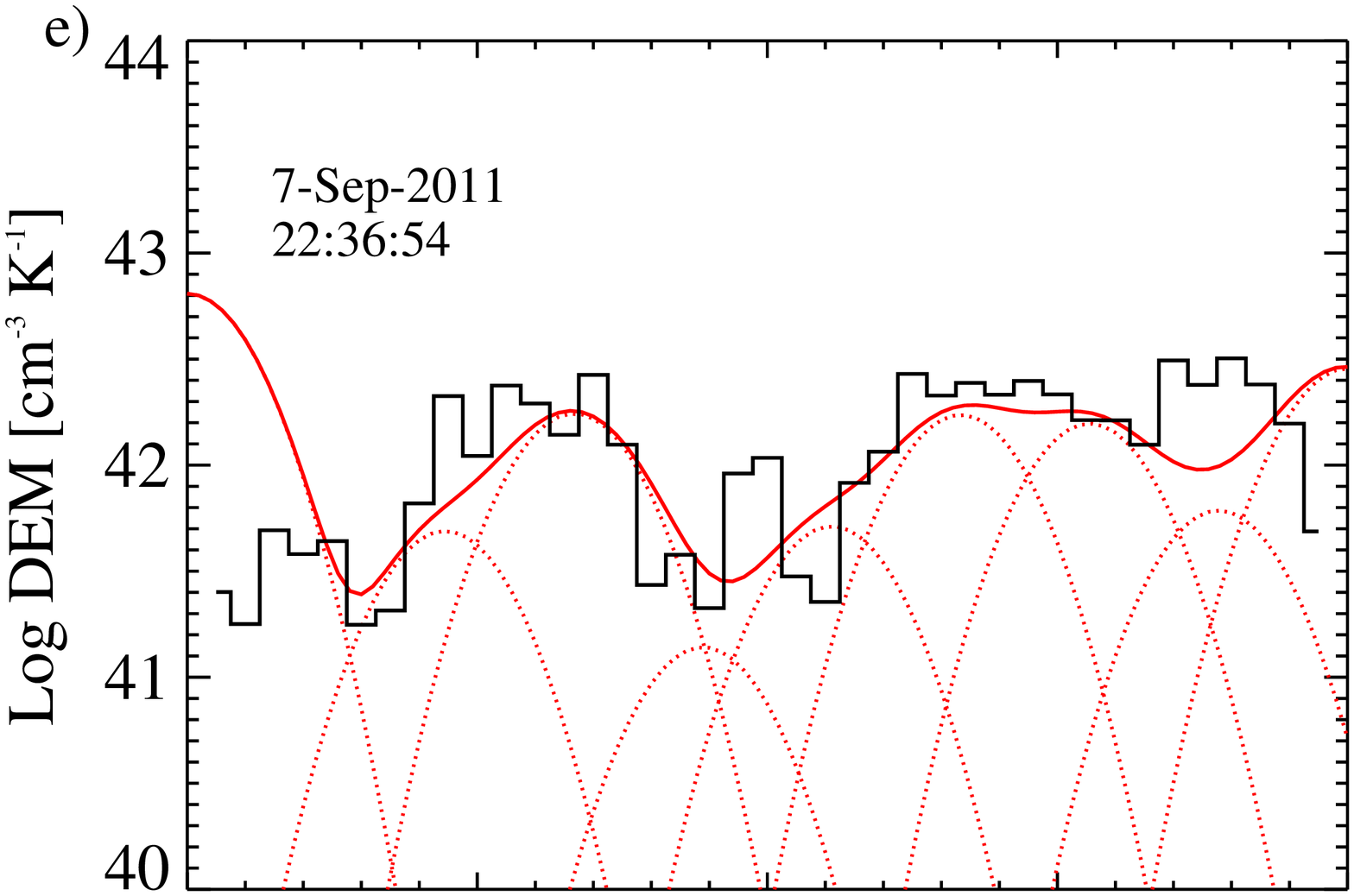} &
\hspace{-0.21in}\includegraphics[width=0.4\linewidth]{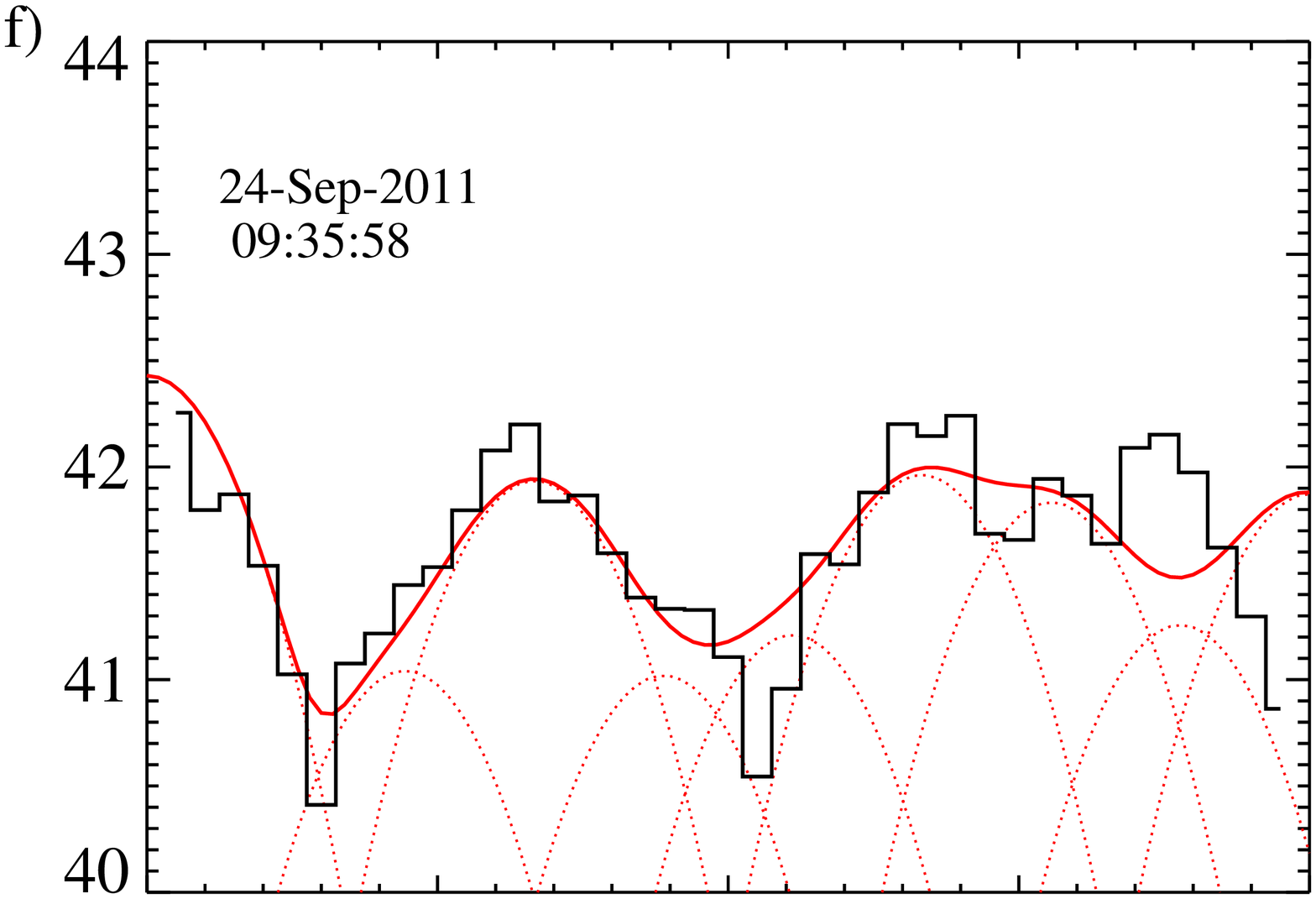} \\
\hspace{-0.04in}\includegraphics[width=0.4\linewidth]{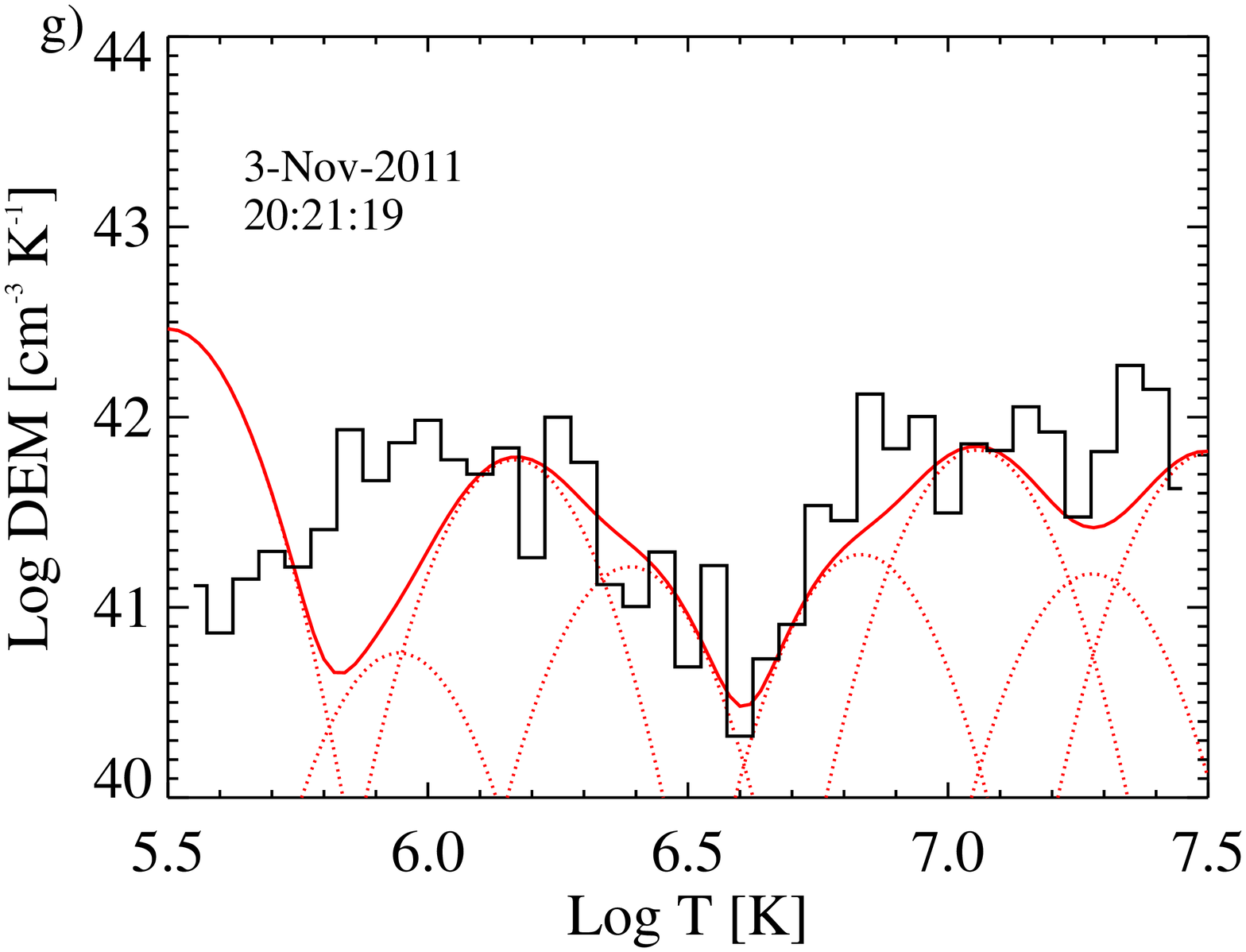} &
\hspace{-0.21in}\includegraphics[width=0.4\linewidth]{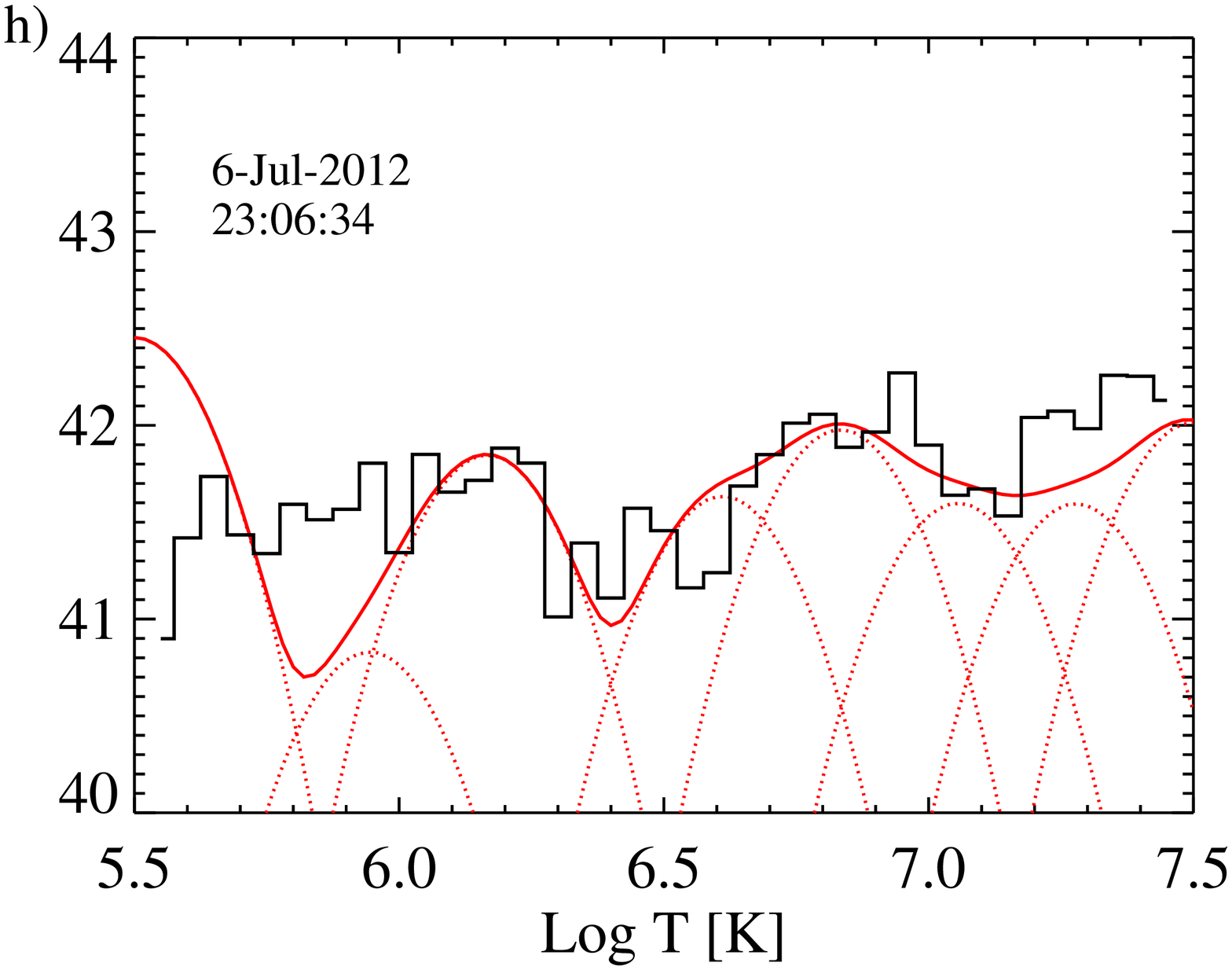} \\
\end{tabular}
\caption{Comparison of DEMs obtained from the MCMC and NRL-EVE construction methods. The black histogram line is the MCMC solution. The red line is the total solution from the \citet{warr13} method and the red dotted lines represent the contributions from the individual Gaussian components.}
\label{f:dem_nrl}
\end{figure}

For eight events the MCMC and NRL-EVE results are shown in Figure \ref{f:dem_nrl}. To directly compare the results of the two methods the output of the MCMC routine is divided by logarithmic temperature to obtain the DEM in units of cm$^{-3}$ K$^{-1}$. The low temperature peak in the distribution is reproduced in the NRL-EVE code and generally good agreement is found between the two different methods, particularly in the case of the 2011 March 09, 2011 August 04 and 2011 September 24 flares (Fig. \ref{f:dem_nrl}, Panel a,c and f respectively).

The small number of unblended emission lines available in the MEGS-A wavelength range limits the analysis as the structure of the solution is not well constrained at low temperature.  Any DEM analysis is sensitive to the adopted line fluxes and uncertainties, and it is possible that the structure in the derived DEMs may change due to updated instrument calibration in the future. The results presented in this study are obtained from the most recent EVE data version, using the most up-to-date atomic data available, an abundance free construction, and verified using a second construction method which gives solutions in good agreement with the MCMC method. 

\subsection{Location of Flare Emission}\label{ss:aia_imag}

\begin{figure*}
\centering
\begin{tabular}{ccc}
\hspace{-0.25in}\includegraphics[width=0.35\linewidth]{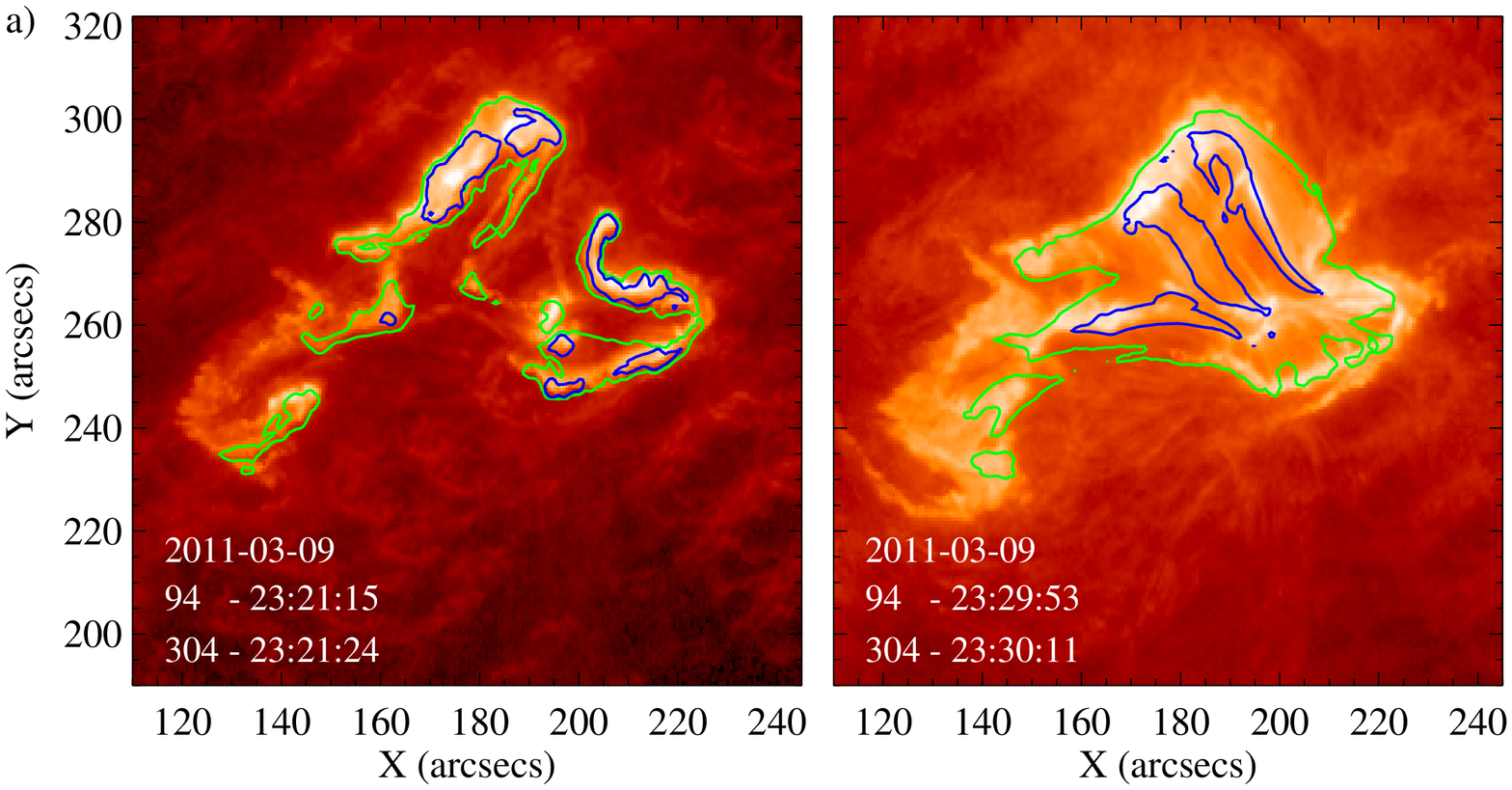} &
\hspace{-0.2in}\includegraphics[width=0.35\linewidth]{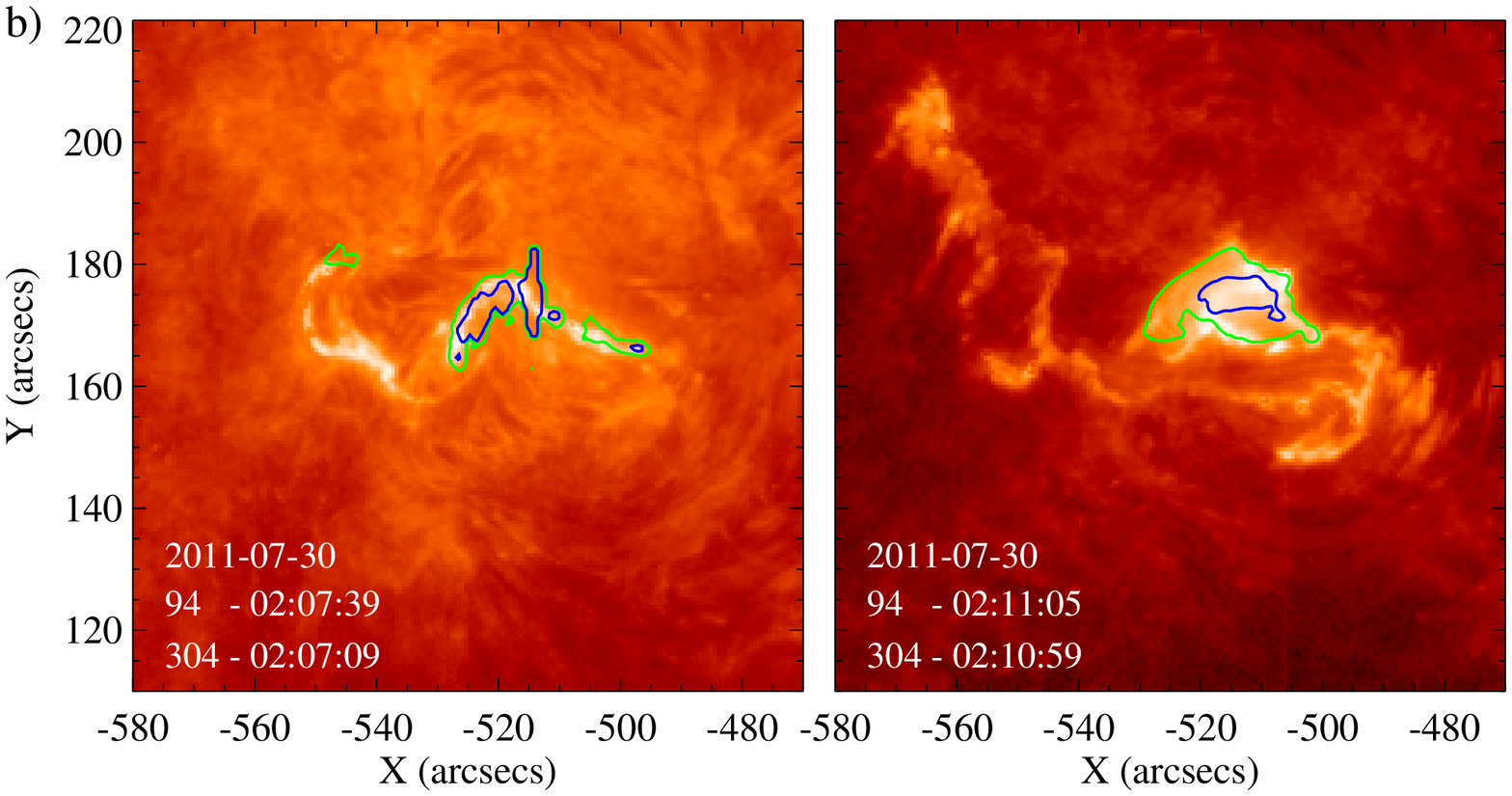} &
\hspace{-0.2in}\includegraphics[width=0.35\linewidth]{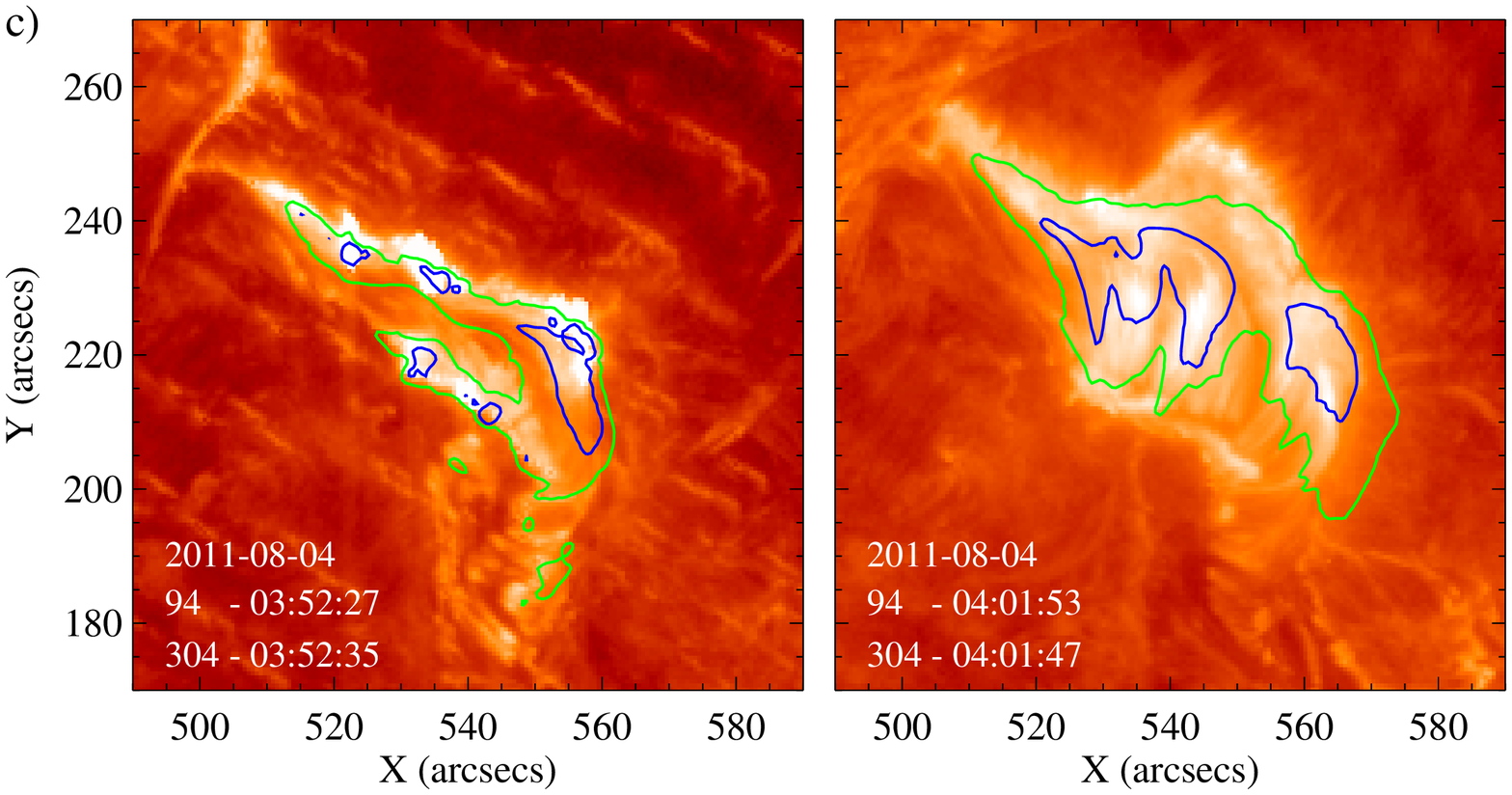} \\
\hspace{-0.25in}\includegraphics[width=0.35\linewidth]{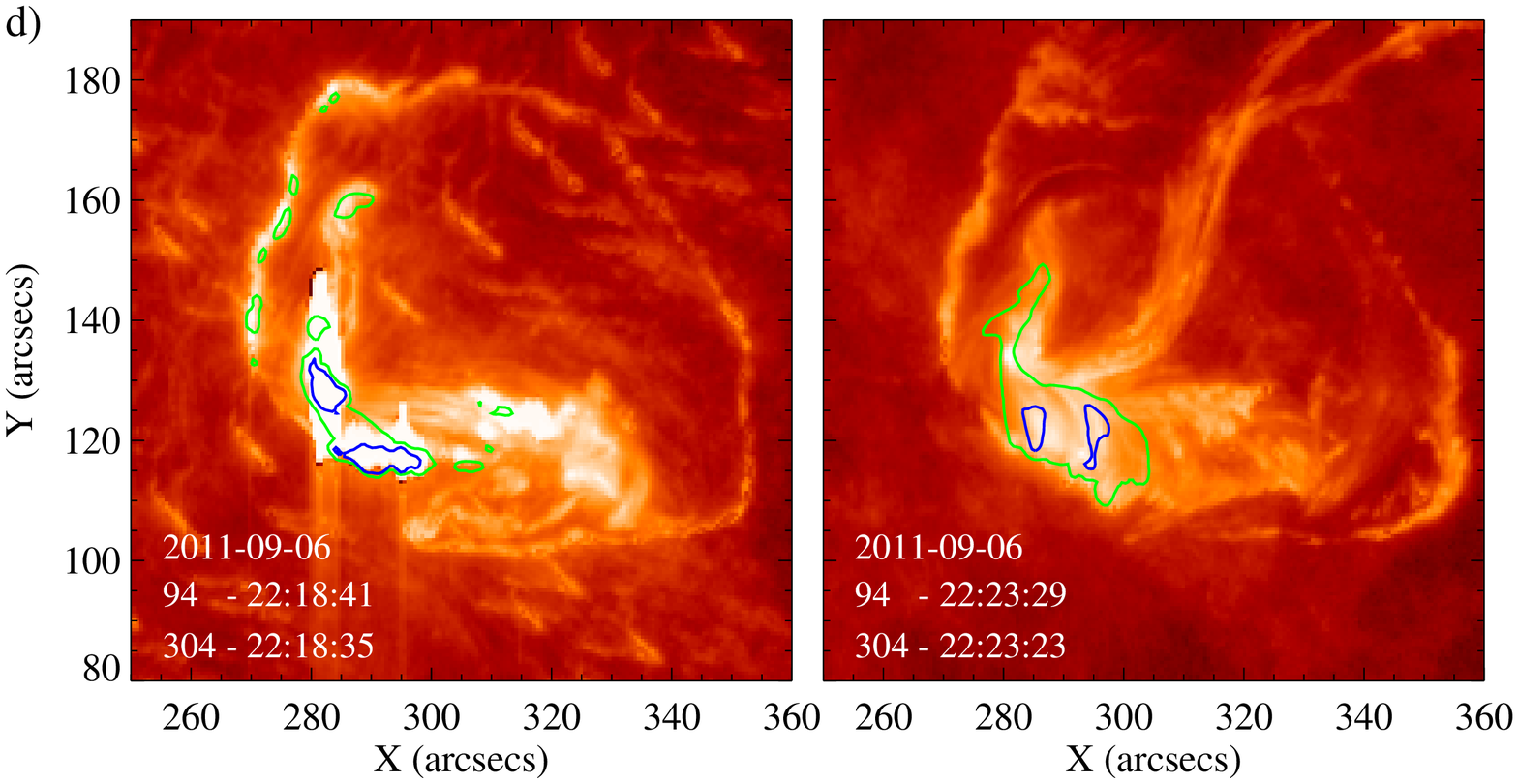} &
\hspace{-0.2in}\includegraphics[width=0.35\linewidth]{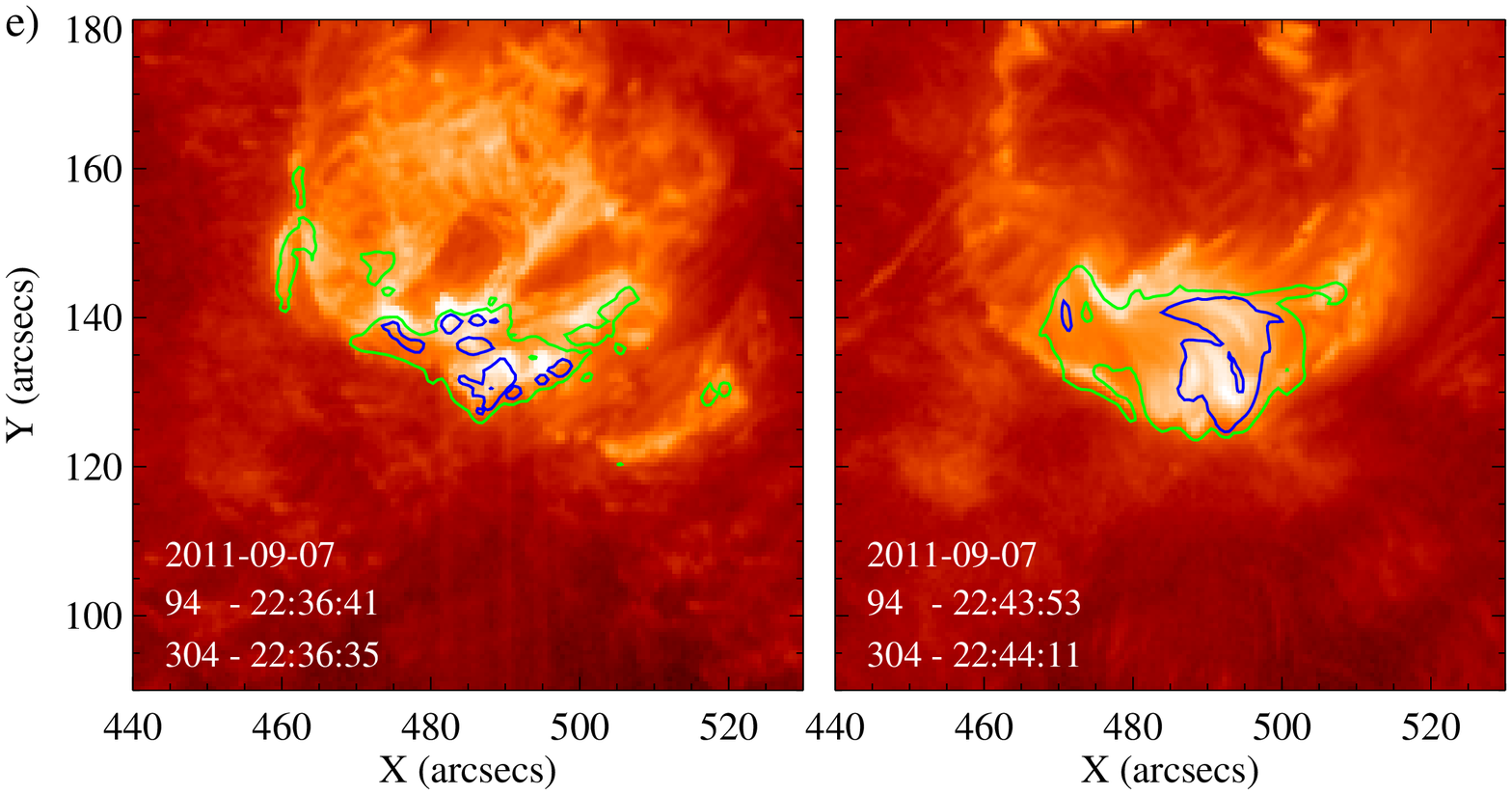} &
\hspace{-0.2in}\includegraphics[width=0.35\linewidth]{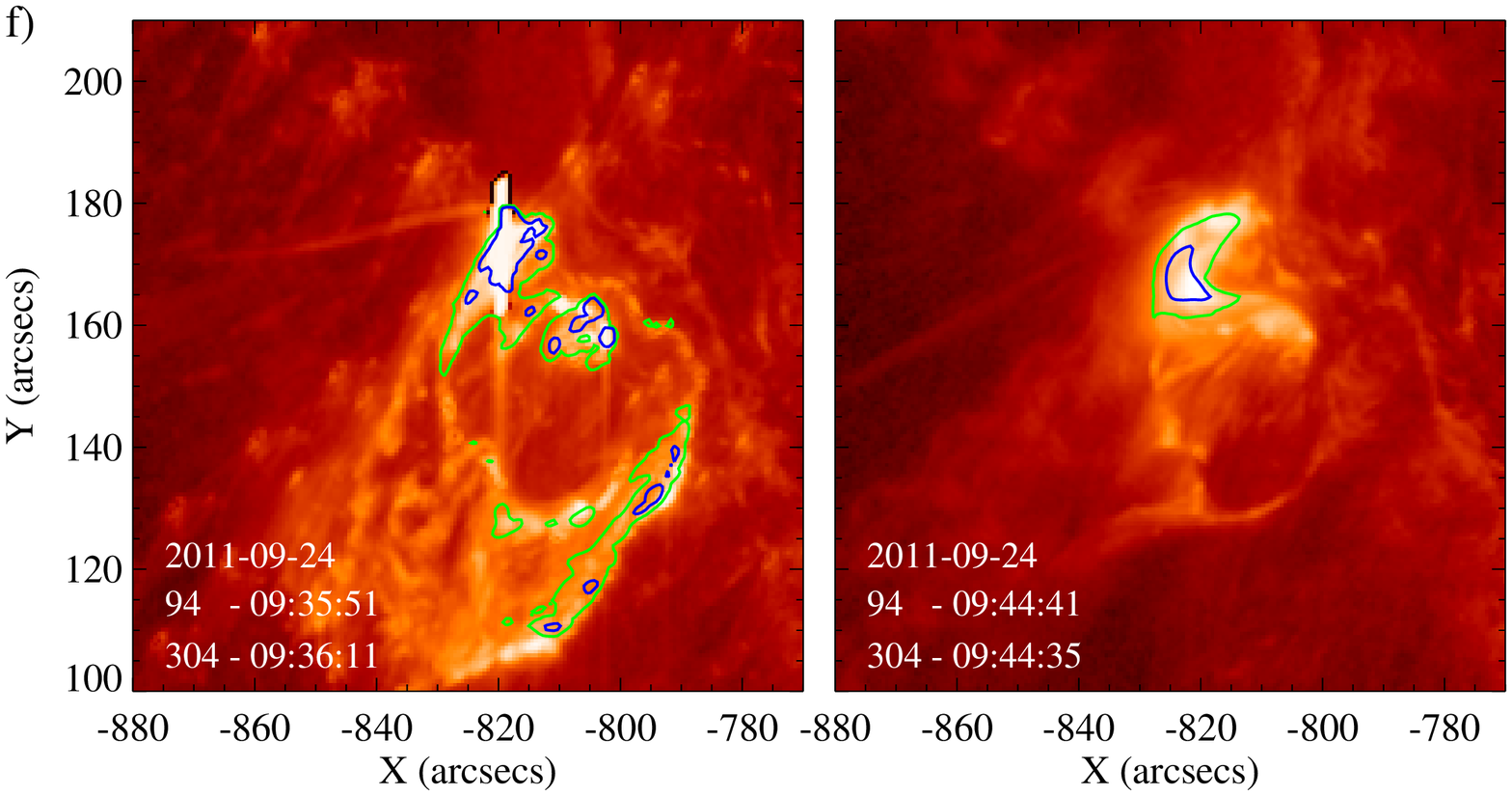} \\
\hspace{-0.25in}\includegraphics[width=0.35\linewidth]{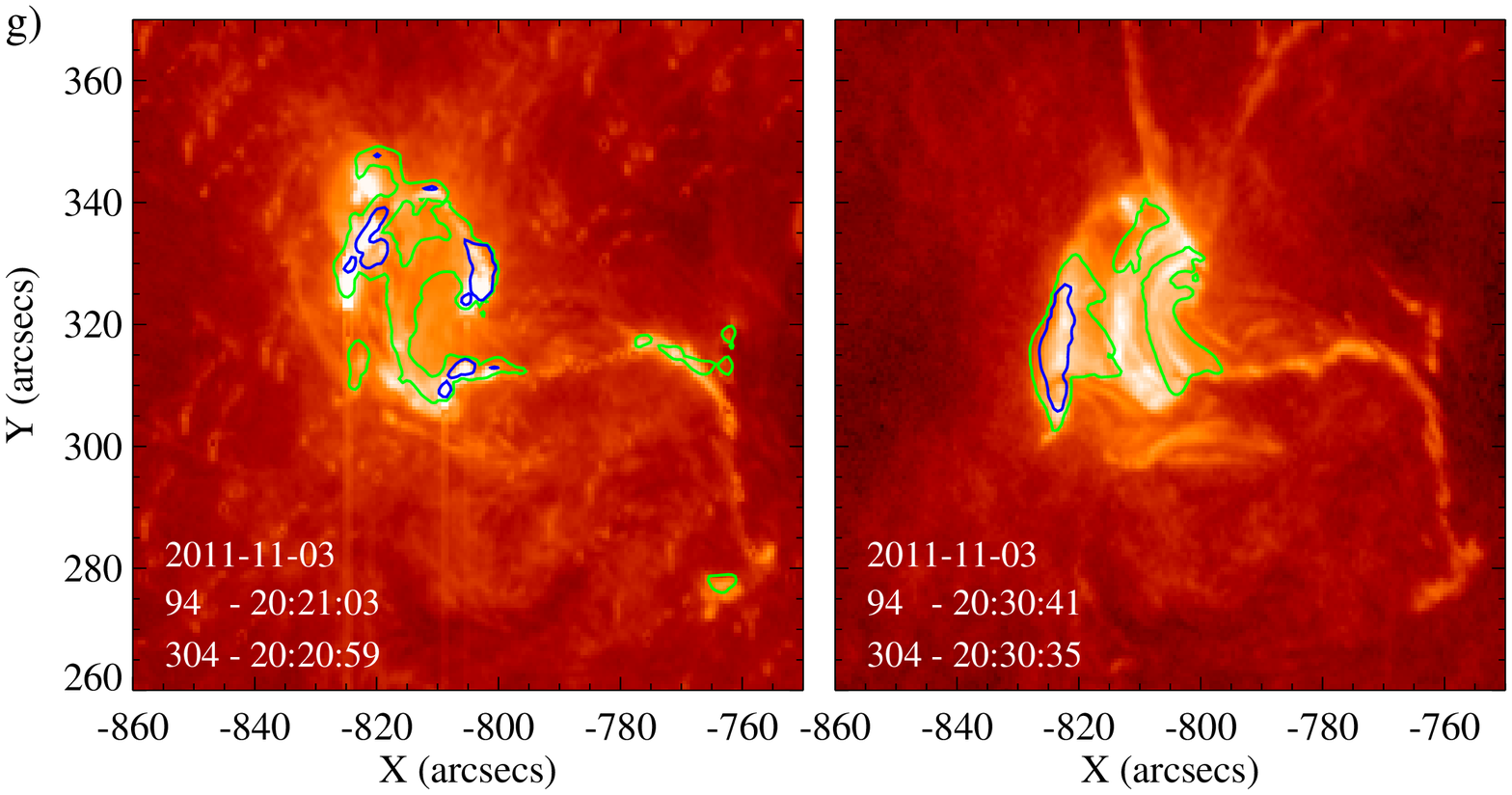} &
\hspace{-0.2in}\includegraphics[width=0.35\linewidth]{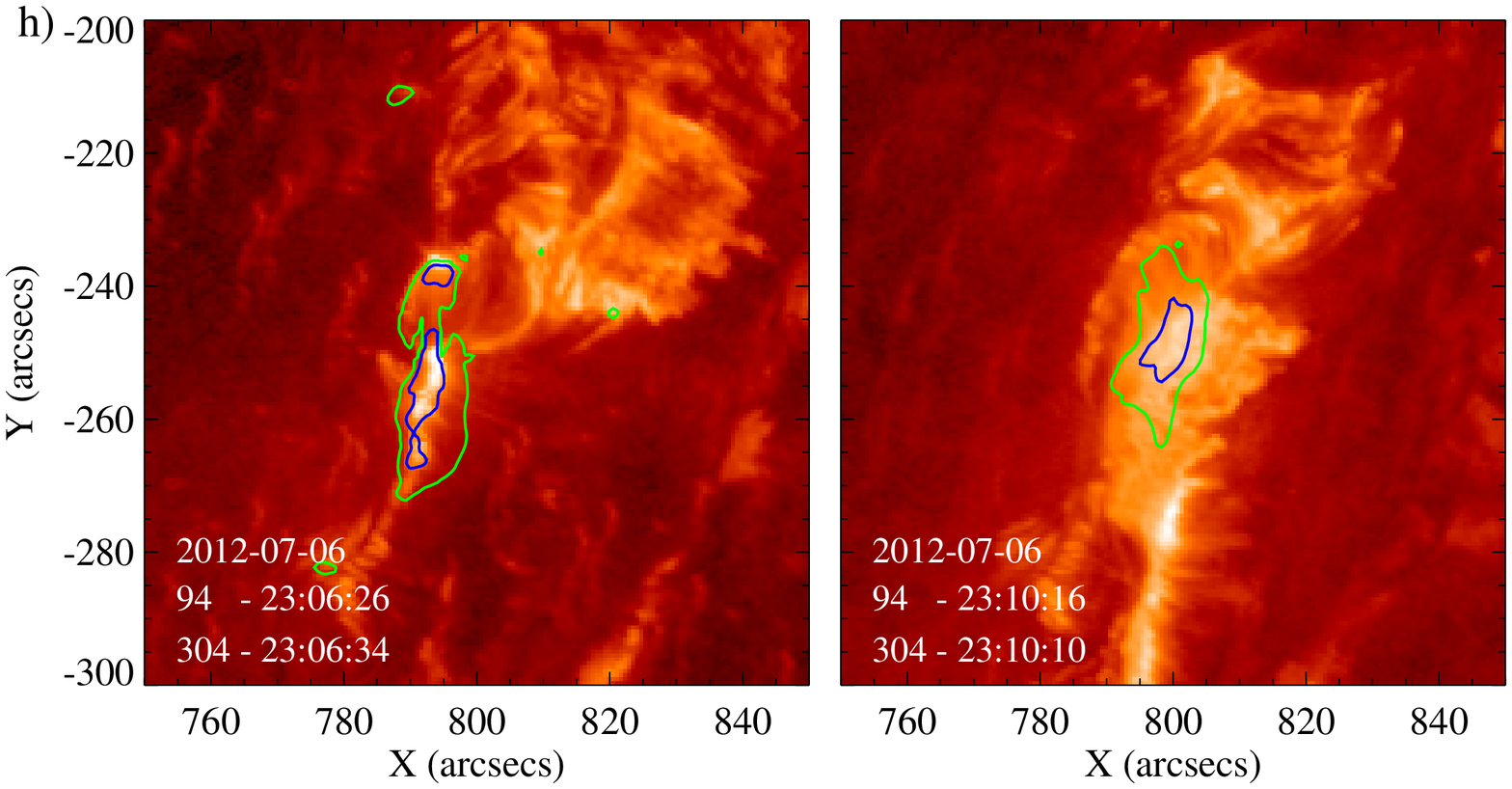} &
\hspace{-0.2in}\includegraphics[width=0.35\linewidth]{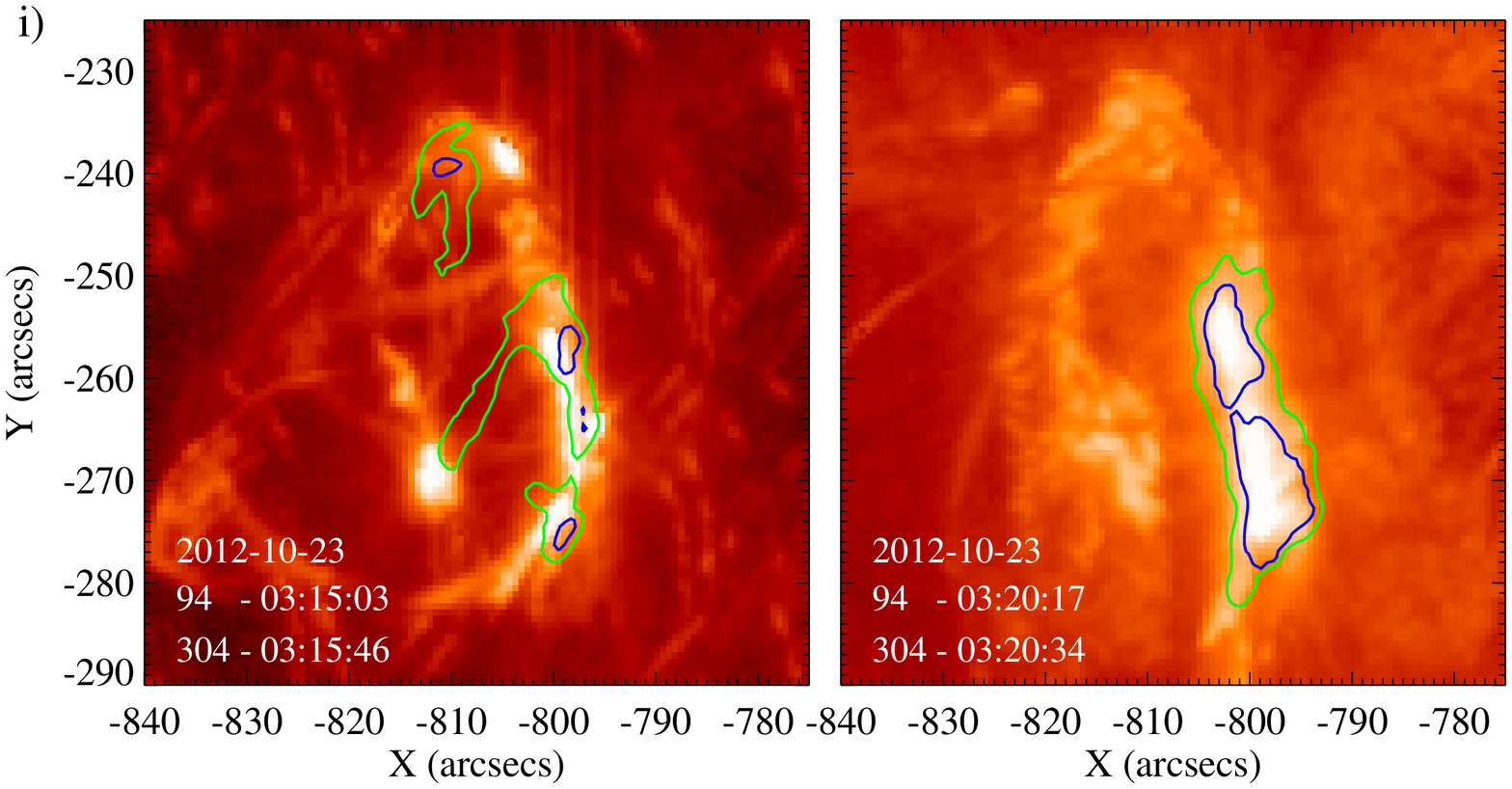} \\
\end{tabular}
\caption{AIA 304\AA\ images from the impulsive phase and decay phase of the events studied. Overlaid are the 10\% and 50\% intensity contours (green and blue respectively) from the 94\AA\ channel. The nearest unsaturated exposures to the time of the DEMs shown in Figure \ref{f:emd_plots} are shown. During the impulsive phase, EUV emission observed by EVE originates from the flare ribbons and footpoints, compared to the loop emission in the decay phase.}
\label{f:aia_plots}
\end{figure*}

Images from SDO/AIA \citep{leme12}, shown in Fig. \ref{f:aia_plots}, were used to provide spatial information about the flare emission, providing a context for where in the atmosphere the EUV emission observed by EVE is originating from at different points in the flare evolution. For each flare a comparison is shown between the impulsive phase (left panels) and the decay phase (right panels). Intensity contours from the 94\AA\ channel (\ion{Fe}{18}, peak response at log T$_{e}$ = 6.9) at the 10\% and 50\% levels are over-plotted on images from the 304\AA\ channel. Images during the impulsive phase are shown from as close as possible to the time of the EVE observations used to construct the DEMs shown in Figure \ref{f:emd_plots}, based on the exposure time to limit the amount of detector saturation. 

At the peak of the impulsive phase the 10\% contours from the 94\AA\ channel outline the flare ribbons with the highest intensity emission originating from several localised regions, i.e. the flare footpoints/ribbons. The decay phase images and contours show the flare loop arcades. This imaging provides evidence to support that, for these events, the high temperature EUV emission during the impulsive phase is originating from the flaring chromosphere, while during the decay phase the emission is from the coronal loops. Despite the spatially unresolved nature of EVE observations, the imaging indicates that the EVE flare spectra can be used to determine the (spatially averaged) thermal structure of the chromospheric flare emissions during the impulsive phase.

\section{Discussion}\label{s:dis}

The DEMs derived from EVE observations suggest that the EUV emission originates from a region with a peak temperature of 1 - 2 MK, with a large decrease in emission from ions above Fe X. In some events there is no Fe XIV emission observable in the EVE flare spectra at the peak of the impulsive emission. This is similar to the EMDs derived from \textit{Skylab} NRL spectroheliograph observations \citep{widi84}. The impulsive EUV source in the \textit{Skylab} study showed increased emission from \ion{O}{6} to \ion{Fe}{14}, with a possible maximum at \ion{Fe}{10} and weak increases from lines above typical quiet Sun coronal temperatures. AIA imaging of the flares in this study supports the idea that the hot emission (8 - 10 MK) observed by EVE is also originating from the flare ribbons and footpoints.

A recent example of impulsive phase DEMs, and the first study of the temperature distribution of plasma at flare footpoints is the study of \citet{grah13} who found a distribution that was constantly increasing in temperature up to log T$_{e}$ = 6.9, with a gradient $\propto$ T$^{1}$. The EVE results presented in this study show significant differences in the structure of the distribution at low temperature but similarly show a peak in the distribution at around log T$_{e}$ = 6.9. It may not be appropriate to directly compare the two sets of results, as the EMDs derived from EIS observations were constructed from the total line-of-sight emission, with no pre-flare emission subtracted.

As part of a study into flare ribbon energetics, \citet{flet13} determined DEMs of footpoints using AIA observations during the early phase of a M1.0 flare. The AIA derived DEMs presented exhibited a peak in the distribution at approximately 1 MK, a decrease at temperatures of 3-6 MK and a high temperature peak at 10 MK. While the line-of-sight quiet coronal emissions will increase the magnitude of the DEM at low temperature, these results are very similar to what is obtained from the spatially unresolved EVE observations. This may support the conclusion drawn from the comparison with the AIA imaging, which is that the DEMs derived in this study represent the spatially averaged thermal structure of the flare ribbon and footpoints.

\section{Conclusions}\label{s:conclu}

This study has focused on determining the DEM of the impulsive phase flare emission using data from the EVE MEGS-A instrument. Using the MCMC method and the NRL - EVE code, impulsive phase DEMs were constructed from log T$_{e}$ = 5.5 - 7.5 for nine M and X-class flares. A similar distribution in the re-constructed DEMs is found for each event studied. There is a low temperature component peaking at 1-2 MK, and a broad high temperature component from 7-30 MK, with a minimum in the solution at 3 - 4 MK. Excellent agreement is found between the two independent methods and the solutions are able to produce line irradiance and synthetic spectra that match with EVE observations. However, the structure in the solution is not well constrained at low temperature due to the lack of unblended emission lines in the EVE MEGS-A wavelength range. Imaging of the flare emission from AIA verifies the location of the emission observed by EVE during the impulsive phase, providing evidence that the DEMs obtained represent the spatially-averaged thermal structure of the flare footpoints and ribbons. 

\acknowledgements
M.B.K. thanks the Northern Ireland Department of Employment and Learning for the award of a PhD studentship. R.O.M. is grateful to the Leverhulme Trust for financial support from grant F/00203/X, and to NASA for LWS/TR\&T grant NNX11AQ53G. We thank the Air Force Office of Scientific Research, Air Force Material Command, USAF for sponsorship under grant number FA8655-09-13085. The authors would like to thank Philip Chamberlin for providing the EVE temperature lightcurve routine and Harry Warren for providing the NRL-EVE DEM routines. CHIANTI is a collaborative project involving George Mason University, the University of Michigan (USA) and the University of Cambridge (UK). This research has made use of NASA's Astrophysics Data System.

\bibliographystyle{apj}

\begin{table}
\caption{Emission lines used to construct the differential emission measure and flare temperature evolution lightcurves.}
\centering
\begin{minipage}{1.0\textwidth}
\begin{tabular}{lrclrc}
\hline
Ion & $\lambda$ [nm] & log T$_{e}$ [K] & Ion & $\lambda$ [nm] & log T$_{e}$ [K] \\
\hline
Fe VIII\footnote{Blend of 13.094 and 13.124}     &  13.1& 5.75 & Ca XVII \footnote{Blend of 6 \ion{O}{5} lines during impulsive phase.} & 19.285  & 6.75 \\
Fe IX      &  17.107  & 5.90 & Fe XVIII   &   9.393  & 6.85\\
Fe X       &  17.453  & 6.05 &  Fe XVIII   &  10.395  & 6.85 \\
Fe X       &  17.724  & 6.05  & Fe XIX     &  10.835  & 6.95 \\
Fe XIV     &  21.132  & 6.30  & Fe XX      &  12.184  & 7.00 \\
Fe XV      &  28.461  & 6.35 & Fe XXI     &  12.875  & 7.05 \\
 Fe XVI     &  33.541& 6.45 & Fe XXII    &  13.579  & 7.10\\
Ni XVII    &  24.918 & 6.45 &  Fe XXIV    &  19.202  & 7.25 \\
 Ni XVIII   &  29.198  & 6.50 &   Fe XXIV    &  25.511  & 7.25 \\
 Ca XVI     &  20.860  & 6.70 &  Ni XXVI    &  16.537  & 7.40 \\
\hline
\end{tabular}
\end{minipage}
\label{t:linelist}
\end{table}

\begin{table}
\caption{GOES classification of flares analysed in this paper.}
\centering
\begin{tabular}{llllll}
\hline
Date & Class & Background & Start & Peak & End \\
\hline
2011 March 09          & X1.5 & 22:40 - 22:45 & 23:13 & 23:23 & 23:29 \\
2011 July 30             & M9.3 & 01:45 - 01:50 & 02:04 & 02:09 & 02:12 \\
2011 August 04        & M9.3 & 03:30 - 03:35 & 03:41 & 03:57 & 04:04 \\
2011 September 06 & X2.1 & 21:00 - 21:05 & 22:12 & 22:20 & 22:24 \\
2011 September 07 & X1.8 & 22:00 - 22:05 & 22:32 & 22:38 & 22:44 \\
2011 September 24 & X1.9 & 09:15 - 09:20 & 09:21 & 09:40 & 09:48 \\
2011 November 03  & X1.5 & 19:58 - 20:03 & 20:16 & 20:27 & 20:32 \\
2012 July 06             & X1.1 & 22:32 - 22:37 & 23:01 & 23:08 & 23:14 \\
2012 October 23      & X1.8  & 02:05 - 02:10 & 03:13 & 03:17 & 03:21 \\
\hline
\end{tabular}
\label{t:flare_list}
\end{table}

\begin{table}
\centering
\caption{Parameters of the low temperature DEM component; the peak temperature and total emission measure of the DEM below log T$_{e}$ = 6.6.}
\begin{tabular}{lccc}
\hline
Event & Time &  Log EM [cm$^{-3}$] & log T$_{e}$ [K] \\
\hline
2011 Mar 09 & 23:21:08 & 47.88 & 6.15   \\
2011 Jul 30 & 02:07:54 & 47.67 & 6.20  \\
2011 Aug 04 & 03:52:55 & 47.74 & 6.15  \\
2011 Sep 06 & 22:18:53 & 47.89 & 6.50   \\
2011 Sep 07 & 22:36:54 & 48.11 & 6.20   \\
2011 Sep 24 & 09:35:58 & 47.78 & 6.15   \\
2011 Nov 03 & 20:21:19 & 47.69 & 6.25  \\
2012 Jul 06 & 23:06:34 & 47.68 & 6.20   \\
2012 Oct 23 & 03:15:51 & 47.70 & 6.25   \\
\hline
\end{tabular}
\label{t:flare_emt}
\end{table}

\end{document}